\documentclass[11pt]{article}
\usepackage{amsfonts,slashed}
\usepackage{arydshln}
\usepackage[greek,english]{babel}
\usepackage{cancel}
\usepackage{upgreek}
\usepackage{float}
\usepackage{url}
\usepackage{latexsym}
\usepackage{amsfonts}
\usepackage{amsmath}
\usepackage{epsfig}
\usepackage{latexsym,amssymb}
\usepackage{amsmath,amssymb,amsthm}
\usepackage{hyperref}
\usepackage{fontenc}
\usepackage{graphicx,tikz}
\usepackage{multirow}
\usepackage{xfrac}
\usepackage[T1]{fontenc}
\usepackage{textcomp}
\usepackage{lmodern}
\usepackage{stmaryrd}
\usepackage{cite}

\usepackage[margin=20pt,small]{caption}
\usepackage{subcaption}

\usepackage{ifpdf}
\ifx\pdfoutput\undefined
   \pdffalse
   \usepackage{cite}
 \else
   \pdfoutput=1
   \pdftrue
\fi
\DeclareFontFamily{OMS}{rsfs}{\skewchar\font'60}
\DeclareFontShape{OMS}{rsfs}{m}{n}{<-5>rsfs5 <5-7>rsfs7 <7->rsfs10 }{}
\DeclareSymbolFont{rsfs}{OMS}{rsfs}{m}{n}
\DeclareSymbolFontAlphabet{\Scr}{rsfs}

\numberwithin{equation}{section}
\def\be{\begin{equation}}
\def\ee{\end{equation}}
\def\ba{\begin{array}}
\def\ea{\end{array}}

\newcommand{\bea}{\begin{eqnarray}}
\newcommand{\eea}{\end{eqnarray}}

\def\={~=~}
\def\*{{}^*}

\def\P{\mathcal{P}}

\def\cD{\mathcal{D}}

\newcommand\smallmath[2]{#1{\raisebox{\dimexpr \fontdimen 22 \textfont 2
      - \fontdimen 22 \scriptscriptfont 2 \relax}{$\scriptscriptstyle #2$}}}

\newcommand\smallotimes{\!\smallmath\mathbin\otimes\!}

\newcommand{\uA}{{\underline{A}}}
\newcommand{\uB}{{\underline{B}}}
\newcommand{\uC}{{\underline{C}}}
\newcommand{\uD}{{\underline{D}}}
\newcommand{\uE}{{\underline{E}}}
\newcommand{\uF}{{\underline{F}}}

\newcommand{\MGrav}{\mathbb{M}_{\rm spin-2}}
\newcommand{\MVec}{\Big(\mathbb{M}_{\rm spin-1}\Big)}
\newcommand{\MScal}{\Big(\mathbb{M}_{\rm spin-0}\Big)}
\newcommand{\MTForm}{\Big(\mathbb{M}_{\rm tensor}\Big)}

\newcommand{\EM}[1]{\textcolor{red}{#1}}
\newcommand{\BL}[1]{\textcolor{blue}{#1}}

\def\={~=~}
\def\*{{}^*}

\def\P{\mathcal{P}}

\def\cD{\mathcal{D}}

\newcommand{\SL}[1]{\mathrm{SL}(#1)}
\newcommand{\SU}[1]{\mathrm{SU}(#1)}
\newcommand{\USp}[1]{\mathrm{USp}(#1)}
\newcommand{\SO}[1]{\mathrm{SO}(#1)}
\newcommand{\En}[1]{\mathrm{E}_{#1(#1)}}
\newcommand{\mbf}[1]{\mathbf{#1}}
\newcommand{\Cas}[1]{{\cal C}(#1)}
\newcommand{\CasT}{{\cal C}_3}
\newcommand{\fl}[1]{{\underline{#1}}}

\newcommand{\gL}{{\cal L}}
\newcommand{\cU}{{\cal U}}
\newcommand{\gM}{{\cal M}}
\newcommand{\cA}{{\cal A}}
\newcommand{\cB}{{\cal B}}
\newcommand{\cF}{{\cal F}}
\newcommand{\cY}{{\cal Y}}

\newcommand{\cP}{{\cal P}}
\newcommand{\vg}{{\mathring{g}}}
\newcommand{\UI}{(U^{-1})}

\begin{document}
\begin{titlepage}
\vfill
\begin{flushright}
HU-EP-22/38
\end{flushright}

\vfill
\begin{center}
	{\LARGE \bf Kaluza-Klein Spectrometry \\[1ex] 
		beyond Consistent Truncations:
		The Squashed $S^7$
	}\\[1cm]
	
	{\large\bf Bastien Duboeuf\,$^{a}{\!}$
		\footnote{\tt bastien.duboeuf@ens-lyon.fr}, Emanuel Malek\,${}^{b}{\!}$
		\footnote{\tt emanuel.malek@physik.hu-berlin.de}, Henning Samtleben\,${}^{a,c}{\!}$
		\footnote{\tt henning.samtleben@ens-lyon.fr} \vskip .8cm}
	
	{\it ${}^a$ ENSL, CNRS, Laboratoire de physique, F-69342 Lyon, France}\\ \ \\
	{\it  $^{b}$ Institut f\"{u}r Physik, Humboldt-Universit\"{a}t zu Berlin,\\
		IRIS Geb\"{a}ude, Zum Gro{\ss}en Windkanal 6, 12489 Berlin, Germany}\\ \ \\
	{\it  $^{c}$ Institut Universitaire de France (IUF)}\\ \ \\
	
\end{center}
%%%%%%%%%%%%%%%%%%%%%%%%%%%%%%%%%%%%%%%%%%%%%%%
\vfill

\begin{center}
	\textbf{Abstract}
	
\end{center}
\begin{quote}
	We show how to use Exceptional Field Theory to compute the full Kaluza-Klein spectra of 10- and 11-dimensional supergravity around deformations of backgrounds of maximal gauged supergravity by scalar modes that do not form part of the consistent truncation. This includes deformations of AdS$_4 \times S^7$ and AdS$_5 \times S^5$ by modes that are not part of the ${\cal N}=8$ supermultiplet. As an application, we compute the full Kaluza-Klein spectrum of the ${\cal N}=1$ and ${\cal N}=0$ squashed $S^7$. In this example, all conformal dimensions are captured by a universal formula in terms of the Casimir operators and additional quantum numbers which organise the spectrum.
\end{quote}
\vfill
\setcounter{footnote}{0}

\end{titlepage}
\tableofcontents \noindent {}
\newpage
\section{Introduction} \label{s:Intro}
Compactifications are ubiquitous in string theory and its low-energy limit, 10-/11-dimensional supergravity. One of the hallmarks of the compactification is the appearance of infinitely many massive fields in the lower-dimensional theory. From the lower-dimensional perspective, these Kaluza-Klein (KK) towers encode the information about the geometry and fluxes of the compactification. The Kaluza-Klein modes also play an important role in applications. For example, for non-supersymmetric backgrounds, the Kaluza-Klein masses determine their perturbative stability. Moreover, in AdS vacua, the Kaluza-Klein masses encode the anomalous conformal dimensions of single-trace operators in the holographically dual CFTs.

Despite their importance, computing the Kaluza-Klein masses is a formidable challenge for most string/supergravity compactifications. Until recently, the full tower of Kaluza-Klein masses could only be computed for coset spaces \cite{Salam:1981xd}. Beyond this, there were some techniques to access subsets of the Kaluza-Klein towers. For example, if the compactification admits a consistent truncation to a lower-dimensional supergravity theory, then the Kaluza-Klein masses of the (finitely many) fields kept in the truncation can be computed in the lower dimensional theory. However, for generic compactifications, only the spin-2 tower could be accessed \cite{Bachas:2011xa}.

A new method for computing all the Kaluza-Klein masses for a large class of string compactification was presented in \cite{Malek:2019eaz,Malek:2020yue}. This method uses the formalism of Exceptional Field Theory (ExFT), a reformulation of 10-/11-dimensional supergravity that unifies the gravitational and flux degrees of freedom, and as a result makes manifest an exceptional symmetry group~\cite{Hohm:2013pua}. Using ExFT, \cite{Malek:2019eaz,Malek:2020yue} showed how to compute the full Kaluza-Klein spectrum of any vacuum that can be uplifted from a consistent truncation to ${\cal N}=8$ gauged supergravity. 
Unlike the traditional Kaluza-Klein technology, which requires solution of the eigenvalue spectrum of various internal Laplace operators acting on tensorial harmonics, together with a diagonalisation of the coupled system of higher-dimensional fluctuations, the ExFT method is exclusively based on the tower of scalar harmonics on the internal manifold. The relevant internal vector and tensor harmonics are implicitly taken care of by combining the scalar harmonics with the twist matrix encoding the underlying consistent truncation. Furthermore, to a large extent, the diagonalisation problem has already been resolved by the covariant formulation of ExFT. As a result, the $D=4$ mass spectrum can be computed separately KK level by KK level.

These techniques, for the first time, have given access fo the full Kaluza-Klein spectrum for warped compactifications with few or no remaining (super\nobreakdash-)symmetries \cite{Malek:2019eaz,Malek:2020yue,Malek:2020mlk,Varela:2020wty,Guarino:2020flh,Cesaro:2020soq,Bobev:2020lsk,Giambrone:2021zvp,Cesaro:2021haf,Cesaro:2021tna,Giambrone:2021wsm,Cesaro:2022mbu} and led to many interesting insights. For supersymmetric AdS vacua, the protected part of the Kaluza-Klein spectrum can be matched with the superconformal index of the CFT, as was done for the Pilch-Warner AdS$_5$ vacuum in \cite{Bobev:2020lsk} and for the $\SU{3} \times \mathrm{U}(1)$-invariant AdS$_4$ vacuum in ${\cal N}=8$ SUGRA in \cite{Malek:2020yue,Kim:2021zlh}. Moreover, the Kaluza-Klein spectrum can be used to determine compactness of the conformal manifold, which may not be visible in the consistent truncation \cite{Giambrone:2021wsm,Cesaro:2022mbu}, see also \cite{Guarino:2021kyp}. For non-supersymmetric vacua that are perturbatively stable within the consistent truncation, \cite{Malek:2020mlk} showed that instabilities can be nonetheless triggered from higher Kaluza-Klein modes. Finally, the Kaluza-Klein spectroscopy can also be used to prove the existence of perturbatively stable non-supersymmetric AdS vacua in 10-dimensional supergravity \cite{Guarino:2020flh,Giambrone:2021wsm}.

Despite these successes, one drawback of the method of \cite{Malek:2019eaz,Malek:2020yue} is that it only applies to vacua that can be uplifted from ${\cal N}=8$ supergravity via a consistent truncation. This means that these vacua arise by deforming round spheres/hyperboloids by the scalar fields that live in the lower-dimensional ${\cal N}=8$ supermultiplet. Yet there are many interesting vacua that arise from deformations by scalar fields that are not part of the ${\cal N}=8$ truncation, such as the AdS$_4 \times$ squashed $S^7$ vacua with ${\cal N}=1$ and ${\cal N}=0$ supersymmetry \cite{Awada:1982pk,Duff:1983ajq}, or AdS vacua obtained by TsT deformations \cite{Lunin:2005jy,Bobev:2021gza,Bobev:2021rtg}.

In this paper, we show that the methods of \cite{Malek:2019eaz,Malek:2020yue} can in fact be extended to compute the full Kaluza-Klein spectrum for vacua that are generic supergravity deformations of backgrounds of ${\cal N}=8$ supergravity, 
i.e.\ where the deformation does not necessarily take place within the ${\cal N}=8$ consistent truncation.\footnote{Strictly speaking, the undeformed background of ${\cal N}=8$ supergravity does not necessarily have to be a solution to the equations of motion.} 
The method applies to deformations that are triggered by the 10- or 11-dimensional supergravity fields, as opposed to stringy deformations (in the AdS/CFT context, we are thus considering single-trace deformations).
The technical reason underlying the success of our method is that such vacua are generalised parallelisable in ExFT, i.e.\ they admit a globally well-defined generalised frame \cite{Lee:2014mla}. This allows us to compute the entire Kaluza-Klein spectrum by only knowing the scalar harmonics of the background, drastically simplifying the computations. However, unlike vacua that form part of the ${\cal N}=8$ truncation, the vacua we are considering here are no longer generalised \textit{Leibniz} parallelisable, i.e.\ their generalised frames no longer form an algebra. This is captured by a non-constant intrinsic torsion tensor and causes level mixing, or ``space-invaders'' \cite{Duff:1986hr}, in the Kaluza-Klein spectrum, compared to the undeformed ${\cal N}=8$ background. In fact, our method can be applied to any generalised parallelisable background, even those which are not linked by a deformation to a generalised Leibniz parallelisable one.

As an example, and a demonstration of the power of this method, we compute the full Kaluza-Klein spectrum for the squashed $S^7$ AdS$_4$ vacua of 11-dimensional supergravity. The squashed $S^7$ is a Freund-Rubin compactification that preserves $\USp{4} \times \SU{2} \subset \SO{8}$ isometries of the round $S^7$ and ${\cal N}=1$ \cite{Awada:1982pk} (or ${\cal N}=0$ in the case of the right-squashed $S^7$ \cite{Duff:1983ajq}) supersymmetry. While it is a coset space, it is not a symmetric space, which has in fact obstructed the determination of the complete Kaluza-Klein spectrum on this background, although major parts of it have been put together over the years \cite{Nilsson:1983ru,Yamagishi:1983ri,Duff:1986hr,Nilsson:2018lof,Ekhammar:2021gsg,Karlsson:2021oxd}.
Our method finally allows to compute the full spectrum around the squashed $S^7$. 
For the ${\cal N}=1$ squashing, states consistently organise into ${\cal N}=1$ supermultiplets and we find that the
conformal dimensions of the superconformal primaries are given by the intriguing universal formula
\begin{equation}
	\Delta = 1 + \frac53 s + \frac13 \sqrt{(3J+2s^2)^2 + 5\, \CasT} \,.
\end{equation}
Here, $J$ is the spin of the primary and $\CasT$ denotes a linear combination of the $\USp{4}$ and $\SU{2}$ Casimirs. Finally, $s\in\frac12\mathbb{Z}$ is an additional $\mathbb{R}^+$ charge that may be introduced to organise the spectrum. Interestingly, for all but a few small representations, we find that the range of $s$ amounts to the $L_z$ eigenvalues of fixed $\SL{2}$ representations, 
suggesting to elevate this $\mathbb{R}^+$ to a full $\SL{2}$ group.

The outline of our paper is as follows. We begin in section \ref{s:Review} with a brief review of the relevant parts of the $\En{7}$ ExFT. In section \ref{s:KKBeyondCT}, we show how to extend the methods of \cite{Malek:2019eaz,Malek:2020yue} and compute the Kaluza-Klein spectrum for vacua that do not form part of a consistent truncation. In section \ref{s:SquashedS7Sect}, we review the squashed $S^7$ and apply the new technology to compute its full Kaluza-Klein spectrum. Finally, we conclude with a discussion of our findings and an outlook in section \ref{s:Conclusions}. We round off our paper with appendix \ref{app:E6}, where we give the mass matrices for five-dimensional compactifications, and appendices \ref{app:Spectrum}, \ref{app:Degeneracies}, which present further details of the squashed $S^7$ spectra.

%%%%%%%%%%%%%%%%%%%%%%%%%%%%%%%%%%%%%%%%%%%%%%%%%%

\section{Review of $\En{7}$ ExFT} \label{s:Review}

%%%%%%%%%%%%%%%%%%%%%%%%%%%%%%%%%%%%%%%%%%%%%%%%%%

In this section, we briefly review the relevant features of $\En{7}$ ExFT here and refer to \cite{Hohm:2013uia} for details. $\En{7}$ ExFT is a reformulation of 10-/11-dimensional supergravity that makes an $\En{7}$-dimensional symmetry manifest by combining the metric and fluxes into larger generalised tensors, which form representations of $\En{7}$. This formulation of higher-dimensional maximal supergravities is tailored to the description of compactifications to four dimensions.
The bosonic field content of the theory consists of
\begin{equation} \label{eq:ExFTfieldsE7}
	\left\{ g_{\mu\nu},\, \gM_{MN},\, {\cal A}_{\mu}{}^M,\, {\cal B}_{\mu\nu\,\alpha},\, {\cal B}_{\mu\nu\,M} \right\} \,,\qquad \mu, \nu = 0, \dots, 3\,, \quad M = 1, \dots, 56 \,, \quad \alpha = 1, \ldots, 133 \,,
\end{equation}
where $g_{\mu\nu}$ is the four-dimensional metric and the generalised metric $\gM_{MN}$ parameterises the coset space $\En{7}/\SU{8}$. 
The indices $M, N, \dots$ and $\alpha, \beta, \dots$ label fundamental and adjoint indices of $\En{7}$, respectively.
All fields \eqref{eq:ExFTfieldsE7} depend on 4 external coordinates $x^\mu$ and internal coordinates $y^m$, chosen such that the embedding of the gradient operators $\partial_m \hookrightarrow \partial_M$ satisfies the covariant section condition
\begin{equation} \label{eq:sectionE7}
	\Omega^{MK}\,(t_\alpha)_K{}^{N}\,\partial_M \Phi_1\, \partial_N \Phi_2  = 0 = \Omega^{MN} \partial_M \Phi_1\, \partial_N \Phi_2 \,,
\end{equation}
where $\Phi_1$ and $\Phi_2$ refer to any two fields of the theory or their products. There are two inequivalent maximal ways to solve \eqref{eq:sectionE7}, with either all fields only depending on 7 internal coordinates or 6 internal coordinates, corresponding to 11-dimensional and type IIB supergravity, respectively.

The gauge structure of  $\En{7}$ ExFT is encoded in the generalised Lie derivative, which for a generalised vector field of weight $\lambda$, is defined as \cite{Coimbra:2011ky,Berman:2012vc}
\begin{equation} \label{eq:gen_diff_M}
	\gL_\Lambda V^M = \Lambda^N \partial_N V^M - 12\, \partial_K \Lambda^L \, \mathbb{P}^K{}_L{}^M{}_N \, V^N + \lambda\, V^M \partial_N \Lambda^N \,,
\end{equation}
in terms of the projector onto the adjoint representation of $\En{7}$
\begin{equation}
	\mathbb{P}^K{}_M{}^L{}_N = (t_\alpha)^K{}_M (t^\alpha)^{L}{}_N = \frac1{24}\,\delta_M^K\delta_N^L + \frac1{12}\,\delta_M^L\delta_N^K + (t_\alpha)_{MN} (t^\alpha)^{KL} - \frac1{24} \,\Omega_{MN} \Omega^{KL} \,,
\end{equation}
where $(t_\alpha)_M{}^{N}$ and $\Omega^{MN}$ denote the 133 generators and the symplectic invariant of $\En{7}$, respectively. Throughout, we will raise and lower the fundamental $\En{7}$ indices  $M, N, \dots$ using the symplectic invariant $\Omega_{MN}$ in a northwest-southeast convention, i.e.\ $V^M=\Omega^{MN} V_N$ and $V_M=V^N\Omega_{NM} $. In particular,
\begin{equation}
\label{eq:nw-se}
	\Omega_{MK} \Omega^{NK} = \delta_M^N \,.
\end{equation}

As can be seen from the field content \eqref{eq:ExFTfieldsE7}, apart from two-forms ${\cal B}_{\mu\nu\,\alpha}$ in the adjoint representation, $\En{7}$ ExFT also contains two-forms ${\cal B}_{\mu\nu\,M}$ in the fundamental $\mbf{56}$. These $\cB_{\mu\nu\,M}$ are {covariantly constrained}, i.e.\ they are subject to constraints analogous to \eqref{eq:sectionE7}
\begin{equation} \label{eq:sectionE7B}
	0 = \Omega^{MK}\,(t_\alpha)_K{}^{N}\,{\cal B}_{\mu\nu\,M} \, \partial_N \Phi = \Omega^{MK}\,(t_\alpha)_K{}^{N}\,{\cal B}_{\mu\nu\,M} {\cal B}_{\rho\sigma\,N} \,.
\end{equation}

The dynamics of $\En{7}$ ExFT is described by a pseudo-action, given by
\begin{equation}\label{eq:ExFTE7}
	{\cal L}_{\rm ExFT} = \sqrt{|g|}\, \Big( \widehat{R} +\frac{1}{48}\,g^{\mu\nu}{\cal D}_{\mu}{\cal M}^{MN}\,{\cal D}_{\nu}{\cal M}_{MN} -\frac{1}{8}\,{\cal M}_{MN}{\cal F}^{\mu\nu M}{\cal F}_{\mu\nu}{}^N +\sqrt{|g|}{}^{-1}{\cal L}_{\rm top} -V(g,{\cal M})\Big) \,.
\end{equation}
Here, the covariant derivative involves the gauge fields ${\cA}_{\mu}{}^M$, i.e.
\begin{equation}\label{eq:covDeriv}
	{\cal D}_{\mu}{\cal M}_{MN} = (\partial_\mu -{\cal L}_{{\cal A}_\mu})\,{\cal M}_{MN} \,,
\end{equation}
with non-abelian field strengths ${\cal F}_{\mu\nu}{}^M$ given by
\begin{equation}\label{eq:F}
	\begin{split}
		{\cal F}_{\mu\nu}{}^M &\equiv 2\, \partial_{[\mu} {\cal A}_{\nu]}{}^M -2\,{\cal A}_{[\mu}{}^K \partial_K {\cal A}_{\nu]}{}^M -\frac1{2}\left(24\, (t_\alpha)^{MK} (t^\alpha)_{NL} -\Omega^{MK}\Omega_{NL}\right)\,{\cal A}_{[\mu}{}^N\,\partial_K {\cal A}_{\nu]}{}^L \\
		& \quad - 12 \,  (t^\alpha)^{MN} \partial_N {\cal B}_{\mu\nu\,\alpha} -\frac12\,\Omega^{MN} {\cal B}_{\mu\nu\,N} \,.
	\end{split}
\end{equation}
The remaining terms in \eqref{eq:ExFTE7} are as follows. The Einstein-Hilbert term involves a modified Ricci scalar $\widehat{R}$ of the external metric $g_{\mu\nu}$, with all partial derivatives replaced by covariant derivatives \eqref{eq:covDeriv}, i.e.\ 
\begin{equation}
\label{eq:Dg_cov}
\partial_\mu g_{\nu\rho} \rightarrow \partial_\mu g_{\nu\rho} - {\cal A}_\mu{}^M \partial_M g_{\nu\rho}\,.
\end{equation}
The topological term can be most conveniently defined via its formal exterior derivative
\begin{equation}\label{eq:LtopE7}
{\rm d}	{\cal L}_{top} \propto  \varepsilon^{\mu\nu\rho\sigma\tau}\, \cF_{\mu\nu}{}^M\, \cD_{\rho} \cF_{\sigma\tau\,M} \,,
\end{equation}
where $\varepsilon^{\mu\nu\rho\sigma\tau}$ is the five-dimensional alternating symbol. Finally, the potential term in \eqref{eq:ExFTE7} is given by
\begin{equation}\label{eq:Potential}
	\begin{split}
		V(g,{\cal M}) &= - \frac{1}{48}{\cal M}^{MN}\partial_M{\cal M}^{KL}\,\partial_N{\cal M}_{KL} +\frac{1}{2} {\cal M}^{MN}\partial_M{\cal M}^{KL}\partial_L{\cal M}_{NK}\\
		&\quad -\frac{1}{2}\,g^{-1}\partial_Mg\,\partial_N{\cal M}^{MN}-\frac{1}{4}  {\cal M}^{MN}g^{-1}\partial_Mg\,g^{-1}\partial_Ng -\frac{1}{4}{\cal M}^{MN}\partial_Mg^{\mu\nu}\partial_N g_{\mu\nu} \,. 
	\end{split} 
\end{equation}

The pseudo-action \eqref{eq:ExFTE7} is supplemented by the twisted self-duality equation
\begin{equation}\label{eq:twistedSD}
	{\cal F}_{\mu\nu}{}^M = -\frac12\,\sqrt{|g|}\,\varepsilon_{\mu\nu\rho\sigma}\,\Omega^{MN} {\cal M}_{NK}\,{\cal F}^{\rho\sigma\,K} \,,
\end{equation}
which ensures that only 28 of the 56 vector fields correspond to independent propagating degrees of freedom.

\section{Kaluza-Klein spectrometry beyond consistent truncations} \label{s:KKBeyondCT}

\subsection{Fluctuation Ansatz}
While \cite{Malek:2019eaz,Malek:2020yue} showed how to obtain the full Kaluza-Klein spectrum around any vacuum of maximal gauged supergravity that arises from a consistent truncation, here we will go further and treat more general deformations of vacua of ${\cal N}=8$ supergravity that take us outside the consistent truncation. Any background within the consistent truncation to ${\cal N}=8$ gauged supergravity is generalised Leibniz parallelisable, which consists of the following two conditions:
\begin{itemize}
	\item \textit{Generalised parallelisable}: The generalised tangent bundle is trivial, i.e.\ there is a globally well-defined generalised frame. Put differently, we can define the globally well-defined $\En{7}$ objects 
	\begin{equation} \label{eq:DefTwist}
		\left\{ U_M{}^{\fl{A}},\, \rho \right\} \,,
	\end{equation}
	where $U_M{}^{\fl{A}}$ is an $\En{7}$-valued matrix, known as the twist matrix, and $\rho$ is a nowhere-vanishing scalar density of weight $-1/2$. Crucially, generalised parallelisability is a topological condition.
	\item \textit{Leibniz}: For a generalised parallelisable space to be generalised \emph{Leibniz} parallelisable requires $U_M{}^{\fl{A}}$ and $\rho$ to additionally satisfy the differential condition
	\begin{equation} \label{eq:GenLeib}
	\gL_{\cU_{{\fl{A}}}} \cU_{\fl{B}}{}^M = X_{\fl{AB}}{}^{\fl{C}}\, \cU_{\fl{C}}{}^M \,,
	\qquad
	\mbox{for}\;\; {\cU}_{\fl{A}}{}^M = \rho^{-1}\, {\UI}_{\fl{A}}{}^M \,,
	\end{equation}
	where the so-called \emph{intrinsic torsion} $X_{\fl{AB}}{}^{\fl{C}}$ must be constant. In this case, there exists a consistent truncation to maximal $D=4$ supergravity, with the reduction ansatz for the higher-dimensional fields \eqref{eq:ExFTfieldsE7} encoded by the twist matrix $U$ and scalar density $\rho$ \cite{Lee:2014mla,Hohm:2014qga}. The constant intrinsic torsion defines the embedding tensor of the maximal gauged supergravity arising from the consistent truncation.
\end{itemize}

Let us now consider a general deformation of a background of ${\cal N}=8$ supergravity, which is not triggered by the 70 scalar fields of the ${\cal N}=8$ supergravity. Thus, the deformed background will no longer be part of the consistent truncation. Nonetheless, because we are considering a continuous deformation and generalised parallelisability is a topological condition, our deformed background is still generalised parallelisable. That is, our background admits a globally well-defined twist matrix $U_M{}^\fl{A}$ and scalar density $\rho$. However, if we compute the analogue \eqref{eq:GenLeib}, we find
\begin{equation} \label{eq:DefX}
	\gL_{\cU_\fl{A}} \cU_{\fl{B}}{}^M = X_{\fl{A}\fl{B}}{}^{\fl{C}}(y)\, \cU_{\fl{C}}{}^M \,,
	\end{equation}
where the intrinsic torsion, $X_{\fl{A}\fl{B}}{}^{\fl{C}}(y)$, is no longer constant. Rather, the dependence of the intrinsic torsion, $X_{\fl{A}\fl{B}}{}^{\fl{C}}(y)$, on the internal coordinates is the obstruction to the deformation being part of the consistent truncation.

Nonetheless, we can still employ the same ideas as in \cite{Malek:2019eaz,Malek:2020yue} to compute the full KK spectrum. In the previous case with constant intrinsic torsion, the $D=4$ mass matrices can be given as algebraic expressions in terms of the intrinsic torsion tensor. In the more general case \eqref{eq:DefX} the $D=4$ mass operators are most conveniently expressed in terms of differential operators 
\begin{equation} \label{eq:FlattenedDerivative}
	\partial_{\fl{A}} = \cU_{\fl{A}}{}^M \partial_M \,,
\end{equation}
in the internal space. While the expression of the lower dimensional mass operator in terms of differential operators on the internal manifold
is of course at the heart of any Kaluza-Klein analysis, 
the $\En7$ covariant formulation \eqref{eq:ExFTE7} together with an appropriate fluctuation ansatz reduces the analysis to a differential 
problem involving exclusively the tower of scalar harmonics. As a result, the computation of the full Kaluza-Klein spectrum can be done in analogy to the spin-2 sector and leads to universal covariant expressions for all $D=4$ mass operators, which can be straightforwardly diagonalised in the concrete examples.

The non-constant intrinsic torsion \eqref{eq:DefX} now generically gives rise to level-mixing, i.e.\ mass eigenstates of the deformed background will come from mixing states amongst different KK levels of the undeformed background. By contrast, when the deformation is caused by the 70 scalars of the consistent truncation, the deformed KK mass eigenstates will be linear combinations of states of the \textit{same} KK level of the undeformed background \cite{Malek:2019eaz,Malek:2020yue}. 

Just as in \cite{Malek:2019eaz,Malek:2020yue}, because our vacuum is generalised parallelisable, we can expand any tensor fluctuations in terms of the generalised frame $U_M{}^{\fl{A}}$ and $\rho$. 
Thus, we can write any deformation of a generalised parallelisable space as
\begin{equation} \label{eq:Fluct}
\begin{split}
g_{\mu\nu}(x,y) &= \rho^{-2}(y) \left( \vg_{\mu\nu}(x) + h_{\mu\nu}(x,y) \right) \,, \\
\cA_\mu{}^M(x,y) &= \rho^{-1}(y)\, \UI_\fl{A}{}^M(y)\, A_\mu{}^{\fl{A}}(x,y) \,, \\
\gM_{MN}(x,y) &= U_M{}^\fl{A}(y)\, U_N{}^\fl{B}(y)\, \gM_{\fl{A}\fl{B}}(x,y) \,,
\end{split}
\end{equation}
where $\vg_{\mu\nu}$ is a given $D=4$ background metric and $h_{\mu\nu}(x,y)$, $A_\mu{}^{\fl{A}}(x,y)$ and $\gM_{\fl{A}\fl{B}}(x,y)$ are now scalars on the internal space, whose $Y$-dependence we can further expand in a complete basis of scalar functions, $\cY_\Sigma(y)$. Moreover, if we are interested in linearised deformations, we can rewrite \eqref{eq:Fluct} as
\begin{equation} \label{eq:KKAnsatz}
\begin{split}
g_{\mu\nu}(x,y) &= \rho^{-2}(y) \Big( \vg_{\mu\nu}(x) + \sum_\Sigma \cY_\Sigma\, h_{\mu\nu}{}^{\Sigma}(x) \Big) \,, \\
\cA_\mu{}^M(x,y) &= \rho^{-1}(y)\, \UI_\fl{A}{}^M(y) \sum_\Sigma \cY_\Sigma(y)\, A_\mu{}^{\fl{A},\Sigma}(x) \,, \\
\gM_{MN}(x,y) &= U_M{}^\fl{A}(y)\, U_N{}^\fl{B}(y) \Big( \delta_{\fl{A}\fl{B}} + \delta_{\fl{A}\fl{C}} \, \cP_{I,\fl{B}}{}^{\fl{C}} \sum_\Sigma \cY_\Sigma(y)\, j^{I,\Sigma}(x) \Big) \,, \\
\end{split}
\end{equation}
where the $\cP_{I,\fl{A}}{}^{\fl{B}}$ represent the non-compact generators of $\mathfrak{e}_{7(7)}$, such that $\delta_{\fl{A}\fl{C}} \, \cP_{I,\fl{B}}{}^{\fl{C}}=\delta_{\fl{B}\fl{C}} \, \cP_{I,\fl{A}}{}^{\fl{C}} $. Finally, we are ignoring the two-forms since these do not contribute to the linearised equations of motion. In the following, we will omit the explicit summation symbol over the harmonics $\Sigma$.

Since $\fl{A}$, $\fl{B}$ are really $\SU{8}$ indices, we have two invariants with which we can contract: the $\En{7}$ symplectic invariant $\Omega_{\fl{A}\fl{B}}$ and the {$\SU{8}$ invariant} $\delta_{\fl{A}\fl{B}}$. In order to keep track which tensor is involved in an index contraction in $\SU{8}$ language, we will use the following conventions. All $\En{7}$ indices, including the flattened $\fl{A}$, $\fl{B}$, are raised and lowered using the $\En{7}$ symplectic invariant $\Omega_{\fl{A}\fl{B}}$ in the conventions of \eqref{eq:nw-se}. Thus,
\begin{equation}
V^{\fl{A}}\, W_{\fl{A}} = V^{\fl{A}}\, W^{\fl{B}}\, \Omega_{\fl{B}\fl{A}} \,.
\end{equation}
On the other hand, we will supress the effect of raising/lowering indices with $\delta_{\fl{A}\fl{B}}$, i.e.\ they will simply be written in the same position. For example, repeated flattened $\fl{A}$, $\fl{B}$ indices at the same position are contracted with $\delta_{\fl{A}\fl{B}}$, i.e.
\begin{equation}
V^{\fl{A}}\, V^{\fl{A}} = V^{\fl{A}}\, V^{\fl{B}}\, \delta_{\fl{A}\fl{B}} \,.
\end{equation}
In particular, with these conventions, the non-compact generators $\cP_{I,\fl{A}}{}^{\fl{B}}$ satisfy
\begin{equation}
\cP_{I,\fl{A}}{}^{\fl{B}} = \cP_{I,\fl{B}}{}^{\fl{A}} \,.
\end{equation}

The benefit of the Kaluza-Klein Ansatz \eqref{eq:KKAnsatz} is that the linearised equations of motion of ExFT obtained from the pseudo-action \eqref{eq:ExFTE7}, \eqref{eq:twistedSD} drastically simplify. As a result, we can read off the Kaluza-Klein mass operators, which are determined by the intrinsic torsion $X_{\fl{AB}}{}^{\fl{C}}(y)$ from \eqref{eq:DefX}, and flattened derivative \eqref{eq:FlattenedDerivative}.

Before we compute the mass operators, let us comment on the applicability of the technique developed here. Allowing for a $Y$-dependent $X$-matrix in \eqref{eq:DefX} looks like it may apply to any background of 10-/11-dimensional supergravity, since any such background can be described by a generalised vielbein, akin to \eqref{eq:DefTwist}. However, for a generic background, the generalised vielbein will not be globally well-defined, but only up to $\SU{8}$ transformations. This is a problem, since the intrinsic torsion defined in \eqref{eq:DefX} is not invariant under such $\SU{8}$ transformations.\footnote{Put differently, the intrinsic torsion \eqref{eq:DefX} is defined with respect the generalised identity structure given by the generalised parallelisation of the well-defined $U_{\fl{A}}{}^M$. For more general backgrounds, we can define an intrinsic torsion with respect to the $G$-structure of the background, but this will have a different algebraic structure than $X_{\fl{A}\fl{B}}{}^{\fl{C}}$.} Moreover, the fluctuations in \eqref{eq:KKAnsatz} would not be globally well-defined and instead require not just scalar harmonics but also tensor harmonics in various $\En{7}$ representations. As a result, the setup here does not readily apply for general backgrounds, but only for those which are \emph{generalised parallelisable}, i.e.\ with a globally well-defined twist matrix. Such generalised parallelisable backgrounds include deformations of backgrounds that can be uplifted from maximal gauged supergravity, even when the deformation does not correspond to one of the 70 scalar fields of the ${\cal N}=8$ gauged supergravity. We will discuss concrete examples below.

\subsection{Quadratic constraints}
By plugging the fluctuation Ansatz \eqref{eq:KKAnsatz} into the linearised equations of motion, we will obtain the mass operators for the full Kaluza-Klein tower. Since the equations of motion are quadratic and involve the generalised Lie derivative, our fluctuation Ansatz \eqref{eq:KKAnsatz} implies that the mass operators will involve $X^2$, $\partial X$, $X\partial$ and $\partial^2 $ terms. This is just like in \cite{Malek:2019eaz,Malek:2020yue} (where the action of $\partial$ on harmonics was explicitly parametrised by a representation matrix ${\cal T}$), with the important difference that we now also obtain $\partial X$ terms, since the intrinsic torsion is no longer constant.

However, the quadratic terms in $X$ and $\partial X$ are not all independent. Rather, some of them are linked by quadratic constraints, generalising the quadratic constraints of gauged supergravity, governing constant $X$. Just as in that case, the quadratic constraints are a consequence of the section condition \eqref{eq:sectionE7} and thus of closure of the algebra of generalised Lie derivatives, i.e.
\begin{equation} \label{eq:QCGen}
	\left[ \gL_{\cU_\fl{A}},\, \gL_{\cU_{\fl{B}}} \right] = \gL_{\llbracket \cU_{\fl{A}},\, \cU_{\fl{B}}\rrbracket} \,,
\end{equation}
where we defined the shorthand
\begin{equation}
	\llbracket \cU_{\fl{A}},\, \cU_{\fl{B}} \rrbracket = \gL_{\cU_{\fl{A}}} \cU_{\fl{B}} \,.
\end{equation}

Applying \eqref{eq:QCGen} to $\rho^{-2}$ give the following relation
\begin{equation}
	\partial_{\fl{C}} X_{\fl{A}\fl{B}}{}^{\fl{C}} = \partial_{[\fl{A}} \vartheta_{\fl{B}]} + X_{\fl{A}\fl{B}}{}^{\fl{C}}\, \vartheta_{\fl{C}} \,,
\end{equation}
where
\begin{equation}
	\vartheta_{\fl{A}} = \rho^2 \gL_{\cU_{\fl{A}}} \rho^{-2} = \frac1{28} X_{\fl{A}\fl{B}}{}^{\fl{B}} \,,
\end{equation}
is the analogue of the trombone tensor of ${\cal N}=8$ gauged supergravity. Most interesting vacua have $\vartheta_{\fl{A}} = 0$, which will be what we restrict ourselves to in the following. Thus, $X_{\fl{A}\fl{B}}{}^{\fl{C}}$ satisfies the quadratic constraint
\begin{equation} \label{eq:QC1}
	\partial_{\fl{C}} X_{\fl{A}\fl{B}}{}^{\fl{C}} = 0 \,.
\end{equation}
On the other hand, applying \eqref{eq:QCGen} to $\cU_{\fl{A}}{}^M$ gives the quadratic constraint
\begin{equation} \label{eq:QC2}
	X_{\fl{A}\fl{C}}{}^{\fl{E}}\, X_{\fl{B}\fl{E}}{}^{\fl{D}} - X_{\fl{B}\fl{C}}{}^{\fl{E}}\, X_{\fl{A}\fl{E}}{}^{\fl{D}} + X_{\fl{A}\fl{B}}{}^{\fl{E}}\, X_{\fl{E}\fl{C}}{}^{\fl{D}} = - 2\,  \partial_{[\fl{A}} X_{\fl{B}]\fl{C}}{}^{\fl{D}} + 12 \,\mathbb{P}_{\fl{F}}{}^{\fl{E}}{}_{\fl{C}}{}^{\fl{D}}\, \partial_{\fl{E}} X_{\fl{A}\fl{B}}{}^{\fl{F}} \,.
\end{equation}
For constant $X$, with the r.h.s.\ vanishing, this reproduces the quadratic constraints of gauged supergravity~\cite{deWit:2007kvg}.
By tracing \eqref{eq:QC2}, we recover \eqref{eq:QC1} as well as the relation
\begin{equation}
	X_{\fl{A}\fl{C}}{}^{\fl{D}}\, X_{\fl{B}\fl{D}}{}^{\fl{C}} - 2\, X_{\fl{C}\fl{A}}{}^{\fl{D}}\, X_{\fl{D}\fl{B}}{}^{\fl{C}} = 0 \,.
\end{equation}

Finally, the section condition \eqref{eq:sectionE7} also implies a quadratic relation linking the flat derivatives $\partial_{\fl{A}}$ and $X_{\fl{A}\fl{B}}{}^{\fl{C}}$. This can be deduced from the closure of the generalised Lie derivative \eqref{eq:QCGen} when acting on scalar functions $f(y)$. Then, we have
\begin{equation} \label{eq:QCddXd}
	2\, \partial_{[\fl{A}} \partial_{\fl{B}]} - X_{\fl{A}\fl{B}}{}^{\fl{C}}\, \partial_{\fl{C}} = 0 \,.
\end{equation}
We will use these relations throughout the following sections in deriving the mass operators.

\subsection{Mass operators}

We are now ready to plug the fluctuation Ansatz \eqref{eq:KKAnsatz} into the linearised equations of motions resulting from \eqref{eq:ExFTE7} and \eqref{eq:twistedSD} to obtain the mass operators. The computation closely follows the one presented in \cite{Malek:2019eaz,Malek:2020yue} with additional contributions arising from internal derivatives acting on the intrinsic torsion which is no longer constant. 

\subsubsection{Spin-2}

We begin with the linearised equations of motion for the spin-2 fields. The mass terms come from the final two terms of the potential \eqref{eq:Potential}, i.e.
\begin{equation}
	{\cal L}_{{\rm mass}, g} = \frac14 \sqrt{|g|} \left( \gM^{MN} \partial_M g^{\mu\nu} \partial_N g_{\mu\nu} + \gM^{MN} g^{-2} \partial_M g \partial_N g \right) \,.
\end{equation}
Upon computing the resulting linearised equations of motion and inserting the spin-2 fluctuation Ansatz \eqref{eq:KKAnsatz}, we obtain 
(upon gauge fixing in the external space)
\begin{equation}
	 \cY^\Sigma\, \Box_x h_{\mu\nu\,\Sigma} = h_{\mu\nu\,\Sigma}\, \MGrav\, \cY^{\Sigma} + \ldots \,,
	 \label{eq:Boxh}
\end{equation}
with $\Box_x$ the 4-dimensional Laplace operator and the mass operator $\MGrav$ given by
\begin{equation} \label{eq:MassMatrixGraviton}
	\MGrav = - \partial_{\fl{A}} \partial_{\fl{A}} \,,
\end{equation}
acting on scalar harmonics $\cY_{\Sigma}$. The ellipses $\ldots$ in \eqref{eq:Boxh} refer to couplings to vector and scalar modes, which accounts for the spin-2 Higgs mechanism and do not affect the spin-2 masses. However, these effects will, of course, need to be taken into account when evaluating the lower spin masses, by gauge fixing the appropriate unphysical fluctuations. For later use, we just note that the minimal couplings \eqref{eq:Dg_cov} give rise to couplings of the type
\begin{equation}
\Pi_{\underline{A}} \partial_{(\mu} {\cal A}_{\nu)}{}^{\underline{A}}
\;,\qquad
\Pi_{\underline{A}} = \partial_{\underline{A}}
\;,
\label{eq:Spin1Goldstones}
\end{equation}
on the r.h.s.\ of \eqref{eq:Boxh}. I.e.\ the operator $\Pi_{\underline{A}}$ singles out the Goldstone vectors responsible for the spin-2 Higgs mechanism.

\subsubsection{Spin-1}
The vector masses arise from the standard Higgs mechanism and thus from the couplings between vectors and scalar fields in the scalar kinetic of \eqref{eq:ExFTE7}. Thus, let us consider the linearised covariant derivatives of the scalar fields
\begin{equation} \label{eq:LinScalarCovDer}
	\cD_{\mu} \gM_{MN} = U_M{}^{\fl{A}}\, U_{N}{}^{\fl{B}} \left[ \cP_{I,\fl{A}}{}^{\fl{B}}\, \partial_\mu j^{I,\Sigma}\, \cY_\Sigma - \left(\cA_\mu \bullet j\right)_{\fl{A}}{}^{\fl{B}} - \cP_{I,\fl{A}}{}^{\fl{B}}\, \Pi^{I}{}_{\fl{C}} \left( \cA_\mu{}^{\fl{C}} \right) \right] \,,
\end{equation}
where we defined 
\begin{equation} \label{eq:LinScalarCovDerOp}
	\begin{split}
		\left(\cA_\mu \bullet j\right)_{\fl{A}}{}^{\fl{B}} &\equiv \cP_{I,\fl{C}}{}^{\fl{D}}\, j^{I,\Sigma}\, U_{\fl{A}}{}^{(M}\, U_{\fl{B}}{}^{N)}\, \gL_{\cA_\mu} \left( U_M{}^{\fl{C}}\, U_N{}{}^{\fl{D}}\, \cY_{\Sigma} \right) \,, \\
		\Pi^I{}_{\fl{A}}\left(\cA_\mu{}^{\fl{A}}\right) &\equiv 2\, \cP^I{}_{\fl{C}}{}^{\fl{B}}\, U_{\fl{C}}{}^M\, \gL_{\cA_\mu} U_{M}{}^{\fl{B}} \\
		&= -2 \left( X_{\fl{A}}{}^I - 12\, \cP^{I}{}_{\fl{A}}{}^{\fl{B}}\, \partial_{\fl{B}} \right) \cA_\mu{}^{\fl{A}} \,, \\
		X_{\fl{A}}{}^I &\equiv X_{\fl{A}\fl{B}}{}^{\fl{C}} \, \cP^{I,\fl{B}}{}_{\fl{C}} \,,
	\end{split}
\end{equation}
and $\cA_\mu{}^{\fl{A}} = \cA_\mu{}^{\fl{A}\Sigma}\, \cY_\Sigma$. 
Coset indices $I, J$ are raised and lowered with the non-compact part of the Cartan-Killing form.
From \eqref{eq:LinScalarCovDer}, \eqref{eq:LinScalarCovDerOp}, we obtain the operator
\begin{equation} \label{eq:ScalarCouplings}
	\begin{split}
		\Pi^{I}{}_{\fl{A}} &= -2 \left( X_{\fl{A}}{}^{I} - 12\, {\cal P}^I{}_{\fl{A}}{}^{\fl{B}}\, \partial_{\fl{B}} \right) \,,
	\end{split}
\end{equation}
which will be responsible for the vector masses. To express the mass operator, we will also need the adjoint operator, $\Pi_{\fl{A}}{}^I$, of \eqref{eq:ScalarCouplings}, defined as
\begin{equation} \label{eq:AdjointDef}
	\int dY j_{I}\, \Pi^{I}{}_{\fl{A}} \left( \cA_\mu{}^{\fl{A}} \right) = \int dY\, \Pi_{\fl{A}}{}^{I} \left( j_{I} \right) \cA_\mu{}^{\fl{A}} \,,
\end{equation}
where $j_I = j_{I\,\Sigma} \cY^\Sigma$. Evaluating \eqref{eq:AdjointDef} explicitly, we get
\begin{equation} \label{eq:ScalarCouplingsAdjoint}
	\Pi_{\fl{A},I} = - 2 \left( X_{\fl{A}I} + 12\, {\cal P}_{I,\fl{A}}{}^{\fl{B}}\, \partial_{\fl{B}} \right) \,.
\end{equation}

From the linearised equation of motion following from \eqref{eq:ExFTE7}, upon usual gauge fixing in the external space
\begin{equation}
	\Box_x \cA_\mu{}^{\fl{A}} = \MVec{}^{\fl{A}}{}_{\fl{B}}\, \cA_\mu{}^{\fl{B}} + \ldots \,,
\end{equation}
we find the mass operator given by the self-adjoint combination
\begin{equation}
	\begin{split}
		\MVec{}^{\fl{A}}{}_{\fl{B}} &= \frac{1}{24}\, \Pi_{\fl{A},I}\, \Pi^I{}_{\fl{B}} \,.
	\end{split}
\end{equation}
Using \eqref{eq:ScalarCouplings} and \eqref{eq:ScalarCouplingsAdjoint}, the mass operator takes the explicit form
\begin{equation} \label{eq:MassMatrixVector}
	\MVec{}^{\fl{A}}{}_{\fl{B}} = \frac{1}{6} X_{\fl{A}}{}^I\, X_{\fl{B}I} + 2\, {\cal P}_{I,\fl{A}}{}^{\fl{C}}\, \partial_{\fl{C}} X_{\fl{B}}{}^{I} + 4\, {\cal P}_{I,[\fl{A}}{}^{\fl{C}} X_{\fl{B}]}{}^{I}\, \partial_{\fl{C}} - 24\, {\cal P}_{I,\fl{A}}{}^{\fl{C}}\, {\cal P}^I{}_{\fl{B}}{}^{\fl{D}}\, \partial_{\fl{C}}\, \partial_{\fl{D}} \,.
\end{equation}
The mass operator \eqref{eq:MassMatrixVector} contains not only the physical spin-1 fields, but also the Goldstone vectors for the massive gravitons, as well as massless magnetic duals to all these. Thus, in evaluating the mass spectrum from \eqref{eq:MassMatrixVector}, care must be taken to remove all these unphysical modes, e.g.\ by proper gauge fixing in the internal space.

However, we can further simplify the structure of the  mass operator \eqref{eq:MassMatrixVector} by shifting the masses assigned to these unphysical modes. In particular, consider the magnetic dual of the mass operator \eqref{eq:MassMatrixVector} given by
\begin{equation} \label{eq:MagneticVectorMassMatrix}
	\begin{split}
		\Big(\hat{\mathbb{M}}_{\rm spin-1}\Big){}^{\fl{A}}{}_{\fl{B}} &\equiv \Omega_{\fl{A}\fl{C}}\, \Omega^{\fl{B}\fl{D}}\, \MVec{}^{\fl{C}}{}_{\fl{D}} \\
		&= \frac16 X^{\fl{B}}{}_I\, X^{\fl{A}I} + 2\, {\cal P}_{I}{}^{\fl{A}\fl{C}}\, \partial_{\fl{C}} X^{\fl{B}I} + 4\, {\cal P}_{I}{}^{\fl{C}[\fl{A}} X^{{\fl{B}]}I}\, \partial_{\fl{C}} - 24\, {\cal P}_{I}{}^{\fl{A}\fl{C}}\, {\cal P}^{I,\fl{B}\fl{D}}\, \partial_{\fl{C}}\, \partial_{\fl{D}} \,,
	\end{split}
\end{equation}
which carries the same eigenvalues as \eqref{eq:MassMatrixVector} while satisfying the orthogonality condition
\begin{equation} \label{eq:em-orthogonal}
\MVec{}^{\fl{A}}{}_{\fl{B}} \, \Big(\hat{\mathbb{M}}_{\rm spin-1}\Big){}^{\fl{B}}{}_{\fl{C}}  = 0
\,.
\end{equation}
Using \eqref{eq:MagneticVectorMassMatrix} and \eqref{eq:Spin1Goldstones}, we can then deduce the relation
\begin{align} 
\MVec{}^{\fl{A}}{}_{\fl{B}} 
		+ \Big(\hat{\mathbb{M}}_{\rm spin-1}\Big){}^{\fl{A}}{}_{\fl{B}}
		&= \Big(\mathbb{M}_{\rm spin-1}^{(0)}\Big){}^{\fl{A}}{}_{\fl{B}} + \left( \mathbb{N}^{\fl{A}}{}_{\fl{B}}{}^{\fl{C}} - \mathbb{N}^{\fl{B}}{}_{\fl{A}}{}^{\fl{C}} \right) \partial_{\fl{C}} + \partial_{\fl{C}} \mathbb{N}^{\fl{A}}{}_{\fl{B}}{}^{\fl{C}} + \delta^{\fl{A}}_{\fl{B}}\, \MGrav \nonumber\\
 \label{eq:MassMatrixVectorClean}
 &\quad  + 2 \left( \Pi_{\fl{A}} \Pi_{\fl{B}} + \Pi^{\fl{A}} \Pi^{\fl{B}} \right) \,,
\end{align} 
with
\begin{equation} 
	\begin{split}
\left(\mathbb{M}_{\rm spin-1}^{(0)}\right){}^{\fl{A}}{}_{\fl{B}} &= \frac16 \left( X_{\fl{A}I}\, X_{\fl{B}}{}^I + X^{\fl{B}}{}_I\, X^{\fl{A}I} \right)
\;,\\
\mathbb{N}^{\fl{A}}{}_{\fl{B}}{}^{\fl{C}} &= 2 \left( \cP_{I,\fl{A}}{}^{\fl{C}}\, X_{\fl{B}}{}^I + \cP_I{}^{\fl{A}\fl{C}} \, X^{\fl{B},I} \right) \,.
	\end{split}
\end{equation}

The second line in \eqref{eq:MassMatrixVectorClean} acts only on the unphysical Goldstone fields, and can thus be ignored when computing the masses of the propagating degrees of freedom. Moreover, relations \eqref{eq:MagneticVectorMassMatrix} and \eqref{eq:em-orthogonal} imply that the first line of \eqref{eq:MassMatrixVectorClean} carries as eigenvalues all the masses of the physical vector fields with an (unphysical)  multiplicity of two, which has to be divided out. Equation \eqref{eq:MassMatrixVectorClean} turns out to be very useful for the concrete computations as in particular the quadratic action on the scalar harmonics is exactly given by the spin-2 mass operator $\MGrav$ \eqref{eq:MassMatrixGraviton}. 

\subsubsection{Spin-0}
The scalar masses arise from the scalar potential \eqref{eq:Potential} in the action \eqref{eq:ExFTE7}. Let us first rewrite the equations of motion coming from the scalar potential \eqref{eq:Potential} in terms of the intrinsic torsion. We find that a vacuum must satisfy the equation
\begin{equation}
\label{eq:eom_scalars}
	0 = \left[ 2\,\partial_{\fl{C}}X_{\fl{C}\fl{A}}{}^{\fl{B}} 	- X_{\fl{A}\fl{C}}{}^{\fl{D}} X_{\fl{B}\fl{D}}{}^{\fl{C}}	- \frac17\left( 	X_{\fl{A}\fl{C}}{}^{\fl{D}} X_{\fl{B}\fl{C}}{}^{\fl{D}} + X_{\fl{C}\fl{A}}{}^{\fl{D}} X_{\fl{C}\fl{B}}{}^{\fl{D}} - X_{\fl{C}\fl{D}}{}^{\fl{A}} X_{\fl{C}\fl{D}}{}^{\fl{B}} \right) \right]
\cP_{I,\fl{A}}{}^{\fl{B}} 	\,.
\end{equation}
When the intrinsic torsion $X_{\fl{A}\fl{B}}{}^{\fl{C}}$ is constant, this precisely matches the variation of the ${\cal N}=8$ gauged supergravity potential.

Similar to the computation presented in \cite{Malek:2019eaz,Malek:2020yue}, we can plug in the fluctuation Ansatz \eqref{eq:KKAnsatz} and compute the variation of the potential \eqref{eq:Potential} with respect to the scalar fluctuations $j^{I}=j^{I,\Sigma} \mathcal{Y}_{\Sigma}$ to obtain the equation of motions 
\begin{equation}
	\Box_x j^{I} = \MScal{}^I{}_J j^{J} +  \ldots \,,
\end{equation}
where we can obtain the explicit form of the mass operator as
\begin{equation} \label{eq:MassMatrixScalar}
	\begin{split}
		\MScal{}^I{}_J &= X_{\uA\uE}{}^{\uF} X_{\uB\uF}{}^{\uE} \, (\cP^I\cP_J)_{\uA}{}^{\uB} \\
		& \quad + \frac17 \left( X_{\uA\uE}{}^{\uF} X_{\uB\uE}{}^{\uF} + X_{\uE\uA}{}^{\uF} X_{\uE\uB}{}^{\uF} + X_{\uE\uF}{}^{\uA} X_{\uE\uF}{}^{\uB}  \right) (\cP^I\cP_J)_{\uA}{}^{\uB} \\
		& \quad + \frac27 \left( X_{\uA\uC}{}^{\uE} X_{\uB\uD}{}^{\uE} - X_{\uA\uE}{}^{\uC} X_{\uB\uE}{}^{\uD} - X_{\uE\uA}{}^{\uC} X_{\uE\uB}{}^{\uD} \right) (\cP^I)_{\uA}{}^{\uB}\, (\cP_J)_{\uC}{}^{\uD} \\
		& \quad - 2\, (\cP_J)_{C}{}^{D} \, \partial_C X_{\fl{D}}{}^I -\big[ \cP^I,\cP_J\big]{}_{\fl{A}}{}^{\fl{B}}\, \partial_{\fl{C}}X_{\fl{CB}}{}^{\fl{A}} \\
		&\quad + 2\, \Big( (\cP^I)_{\fl{A}}{}^{\fl{B}} X_{\fl{A}J} - (\cP_J)_{\fl{A}}{}^{\fl{B}} X_{\fl{A}}{}^{I} \Big) \, \partial_B - 2\, \big[ \cP^I,\cP_J\big]{}_{\fl{A}}{}^{\fl{B}} 
		\,X_{\fl{CB}}{}^{\fl{A}} \, \partial_{\fl{C}} \\
		& \quad -\delta^I_J\,\partial_{\fl{A}} \partial_{\fl{A}} + 24\, (\cP^I\cP_J)_{\fl{A}}{}^{\fl{B}}\, \partial_{\fl{B}} \partial_{\fl{A}} \,.
	\end{split}
\end{equation}
It is straightforward to verify that \eqref{eq:MassMatrixScalar} is self-adjoint and thus has real eigenvalues. Once again, this operator yields mass eigenvalues not just for the physical scalars but also for the Goldstone scalars that are eaten by the massive spin-1 and spin-2 fields. Since we are not interested in these unphysical fields, and can gauge fix them away, we are free to shift their mass eigenvalues in a way that simplifies the structure of \eqref{eq:MassMatrixScalar}. Thanks to the Higgs mechanism, the operators \eqref{eq:ScalarCouplings}, \eqref{eq:ScalarCouplingsAdjoint} provide us with projection matrices onto the Goldstone scalars, which we can therefore use to add to \eqref{eq:MassMatrixVector} terms of the form $\Pi^{I}{}_{\fl{A}}\, \Pi^{\fl{A}}{}_J$ which only affect the eigenvalues of the non-physical Goldstone modes.

This allows us to rewrite \eqref{eq:MassMatrixScalar} as
\begin{equation} \label{eq:MassMatrixScalarClean}
	\begin{split}
		\MScal{}^I{}_J &= \Big(\mathbb{M}_{\rm spin-0}^{(0)}\Big){}^{I}{}_{J} + \left( \mathbb{N}^{I}{}_{J}{}^{\fl{C}} - \mathbb{N}_J{}^I{}^{\fl{C}} \right) \partial_{\fl{C}} + \partial_{\fl{C}} \mathbb{N}^{I}{}_{J}{}^{\fl{C}} + \delta^{I}_{J}\, \MGrav \\
		&\quad - \frac{1}{24} \Pi^I{}_{\fl{A}}\, \Pi^{\fl{A}}{}_J \,,
	\end{split}
\end{equation}
which is easily verified to be self-adjoint. The operators in \eqref{eq:MassMatrixScalarClean} consist of
\begin{equation}
	\begin{split}
		\Big(\mathbb{M}_{\rm spin-0}^{(0)}\Big){}^{I}{}_{J} &= X_{\uA\uE}{}^{\uF} X_{\uB\uF}{}^{\uE} \, (\cP^I \cP_J)_{\uA}{}^{\uB} \\
		& \quad + \frac17 \left( X_{\uA\uE}{}^{\uF} X_{\uB\uE}{}^{\uF} + X_{\uE\uA}{}^{\uF} X_{\uE\uB}{}^{\uF} + X_{\uE\uF}{}^{\uA} X_{\uE\uF}{}^{\uB}  \right) (\P^I \cP_J)_{\uA}{}^{\uB} \\
		& \quad + \frac27 \left( X_{\uA\uC}{}^{\uE} X_{\uB\uD}{}^{\uE} - X_{\uA\uE}{}^{\uC} X_{\uB\uE}{}^{\uD} - X_{\uE\uA}{}^{\uC} X_{\uE\uB}{}^{\uD} \right) (\cP^I)_{\uA}{}^{\uB}\, (\cP_J)_{\uC}{}^{\uD} \\
		& \quad + \frac16 \, X_{\fl{A}}{}^{I}\, X_{\fl{A},J} \,,
	\end{split}
\end{equation}
which is quadratic in the intrinsic torsion $X_{\fl{A}\fl{B}}{}^{\fl{C}}$ and does not act on the scalar harmonics, and the combination
\begin{equation}
	\begin{split}
		\mathbb{N}^I{}_J{}^{\fl{C}} &= - 2 X_{\fl{A}}{}^I \cP_{J,\fl{C}}{}^{\fl{A}}- 2 X_{\fl{A}J} \cP^I{}_{\fl{C}}{}^{\fl{A}} - [ \cP^I,\, \cP_J ]_{\fl{A}}{}^{\fl{B}} \left( X_{\fl{C}\fl{B}}{}^{\fl{A}} + \frac72 X_{\fl{A}\fl{B}}{}^{\fl{C}} \right) \,,
	\end{split}
\end{equation}
which multiplies a linear differential operator on the scalar harmonics. Just as in the vector mass matrix \eqref{eq:MassMatrixVectorClean}, the quadratic differential operator on the scalar harmonics in \eqref{eq:MassMatrixScalarClean} is simply given by the graviton mass operator \eqref{eq:MassMatrixGraviton}. For the case of constant intrinsic torsion, the formula \eqref{eq:MassMatrixScalarClean} consistently reduces to the expression derived in~\cite{Guarino:2020flh}.

\section{KK spectrum of squashed $S^7$} \label{s:SquashedS7Sect}

As an application of the presented methods, we will now apply the mass formulas to compute the Kaluza-Klein spectrum of the squashed $S^7$ in 11-dimensional supergravity.
The sphere $S^7$ admits two Einstein metrics: the round metric with $\SO{8}$ isometry, and the ``squashed metric'' which only preserves $\USp{4} \times \SU{2} \subset \SO{8}$ isometry. 
These give rise to two supersymmetric Freund-Rubin AdS$_4 \times S^7$ vacua of 11-dimensional supergravity: the ${\cal N}=8$ vacuum, when the $S^7$ is round and an ${\cal N}=1$ vacuum for the squashed $S^7$~\cite{Awada:1982pk}.
For the squashed $S^7$, the isometry group $\USp{4} \times \SU{2}$ is embedded into $\SO{8}$ such that
\begin{equation} \label{eq:USp4SU2Embedding}
	\mathbf{8}_v \rightarrow \mathbf{\left(4,2\right)} \,, \qquad \mathbf{8}_s \rightarrow \mathbf{\left(4,2\right)} \,, \qquad \mathbf{8}_c \rightarrow \mathbf{\left(5,1\right)} \oplus \mathbf{\left(1,3\right)} \,,
\end{equation}
often also referred to as the left-squashed $S^7$.\footnote{Here, we use standard triality conventions, in which the ${\cal N}=8$ gravitinos transform in the $\mathbf{8}_s$ and the round $S^7$ sphere harmonics in symmetric tensor products of the $\mathbf{8}_v$.} 
Note that there are two other embeddings of $\USp{4} \times \SU{2} \subset \SO{8}$, related to \eqref{eq:USp4SU2Embedding} by triality. The embedding 
\begin{equation} \label{eq:RightSquashedEmbedding}
	\mathbf{8}_s \rightarrow \mathbf{\left(5,1\right)} \oplus \mathbf{\left(1,3\right)} \,,
\end{equation}
gives rise to the right-squashed $S^7$, with the same metric as \eqref{eq:USp4SU2Embedding}, but different sign of flux, yielding a non-supersymmetric AdS$_4$ vacuum \cite{Duff:1983ajq}. Finally, the embedding
\begin{equation} \label{eq:VectorEmbedding}
	\mathbf{8}_v \rightarrow \mathbf{\left(5,1\right)} \oplus \mathbf{\left(1,3\right)} \,,
\end{equation}
does not give rise to an Einstein space, and hence no AdS$_4$ vacuum. Here we will mostly focus on the supersymmetric, left-squashed AdS$_4 \times S^7$ vacuum \eqref{eq:USp4SU2Embedding}, but the results also allow to fully determine the non-supersymmetric spectrum on the right-squashed $S^7$.

\subsection{The squashed $S^7$ in ExFT}

The round $S^7$ has already been extensively studied in the ExFT framework. It is a generalised Leibniz parallelisable background, whose twist matrix $U$ consists of an $\SL{8} \subset \En{7}$ matrix \cite{Lee:2014mla,Hohm:2014qga}. As a Freund-Rubin solution, the generalised vielbein of the $S^7$ solution lives on the coset space
\begin{equation}
\frac{\SL{8}}{\SO{8}} \subset \frac{\En{7}}{\SU{8}}\,,
\label{eq:FR-SL8}
\end{equation} 
which contains precisely the right degrees of freedom to capture a 7-dimensional internal metric and 6-form potential. Explicitly, a general Freund-Rubin solution is described a generalised vielbein of the form
\begin{equation}
{\cal V}_{\rm FR} = {\rm exp}[-6\,\alpha\, \mathring{\omega}\,\zeta^n T_n ]
\begin{pmatrix} \mathring{\omega}^{3/4} & 0\\ 0 & \mathring{\omega}^{-1/4} \, \mathring{e}_m{}^i \end{pmatrix} 
\in {\rm SL}(8)
\;,
\label{eq:FRV}
\end{equation}
where $\mathring{e}_m{}^i$ is the vielbein on the internal seven-dimensional space, $\mathring\omega \equiv {\rm det} \,\mathring{e}_m{}^i$, 
and $\zeta^n$ is a vector field with $\mathring\nabla_n\zeta^n =1$\,. The $T_n$ are the generators which extend $\mathfrak{gl}(7)$ to $\mathfrak{sl}(8)$, and $\alpha$ is a constant related to the seven-form flux of the solution. In our conventions, the round $S^7$ solution corresponds to a sphere of radius 1, and $\alpha=1$\,. Upon embedding $\SL{8} \hookrightarrow \En{7}$, the generalised vielbein is related to the generalised metric of \eqref{eq:ExFTE7} as ${\cal M}={\cal V} {\cal V}^T$\,. The twist matrix describing the consistent truncation around the $S^7$ background is explicitly given by \cite{Lee:2014mla,Hohm:2014qga}
\begin{equation}
	\mathring{U}_{\underline{m}}{}^{a}({\cal Y}) =
	\begin{pmatrix}
	\mathring\omega^{3/4}\left({\cal Y}^a 
-6\,\alpha\,\zeta^n \partial_n {\cal Y}^a \right) \\
	 \mathring\omega^{-1/4}\,\partial_m {\cal Y}^a
	\end{pmatrix}  \Bigg|_{\alpha=1} \in {\rm SL}(8)
	\,,\qquad
	\underline{m}=\{0,m\}\,, \quad a = \{1, \ldots, 8\} \,,
	\label{eq:Uround_alpha}
\end{equation}
in terms of the fundamental sphere harmonics ${\cal Y}^a{\cal Y}^a=1$. It differs from \eqref{eq:FRV} by an ${\rm SO}(8)$ rotation from the right, such that consistently $\mathring{U} \mathring{U}^T ={\cal M}={\cal V} {\cal V}^T$\,. The scale factor $\rho$ is given by $\rho=\mathring\omega^{-1/2}$\,. This twist matrix satisfies \eqref{eq:GenLeib} with constant intrinsic torsion.

We will now give a similar ExFT description of the squashed $S^7$. First of all, since the topology is the same as the round $S^7$, the squashed $S^7$ is also a generalised parallelisable background, i.e.\ it can be described by a globally defined $\En{7}$ twist matrix. Moreover, since the squashed $S^7$ is a Freund-Rubin vacuum, by the argument above, the twist matrix should again be an $\SL{8} \subset \En{7}$ element. Finally, the twist matrix must be a continuous deformation of the one corresponding to the round $S^7$, and the deformation must preserve $\USp{4} \times \SU{2}$, see \eqref{eq:USp4SU2Embedding}.

Let us thus consider the decomposition $\En{7} \rightarrow  \SL{5} \times \SL{3} \times \mathbb{R}^+$, such that the  isometry of the squashed $S^7$ is embedded as the compact subgroup $\USp{4} \times \SU{2}\subset \SL{5} \times \SL{3}$. Under this decomposition, the $\En{7}$ adjoint representation branches as follows
\begin{eqnarray} \label{eq:E7Decomp}
+3&    \mathbf{(5,1)} & [0,1,0]
\nonumber\\
+2&  \mathbf{(\overline{5},\overline{3})} & [0,1,2]
\nonumber\\
+1&  \mathbf{(10,3)} & [2,0,2]
\nonumber\\
0& \mathbf{(1,1)} \oplus \mathfrak{sl}(5) \oplus \mathfrak{sl}(3) 
&
[0,0,0]\oplus
[2,0,0]\oplus
[0,2,0]\oplus
[0,0,2]\oplus
[0,0,4] \,,
\\
-1& \mathbf{(\overline{10},\overline{3})}  & [2,0,2]
\nonumber\\
-2& \mathbf{(5,3)} & [0,1,2]
\nonumber\\
-3& \mathbf{(\overline{5},1)} & [0,1,0]  
\nonumber
\end{eqnarray}
where the vertical axis labels the $\mathbb{R}^+$ charge of the representations. 
The last column collects the $\USp{4} \times \SU{2}$ representation content described by their Dynkin labels.
To construct $\USp{4} \times \SU{2}$-invariant deformations, we consider linear combinations of the $\En{7}$ generators which depend on the $S^7$ coordinates
\begin{equation}
	c(y)_\alpha \, t^\alpha = \sum_\Sigma  c_{\alpha,\Sigma}\,\cY^\Sigma \, t^\alpha \;,
	\label{eq:ctY}
\end{equation}
with the scalar harmonics $\cY^\Sigma$ on the round $S^7$. These harmonics combine into the tower of representations
\begin{equation}
\sum_n [n,0,0,0]_{{\rm SO}(8)} \rightarrow \sum_{n,q} [n-2q,q,n-2q]_{\USp{4} \times \SU{2}}
\;,
\label{eq:tower_harmonics}
\end{equation}
under ${\rm SO}(8)$ and  $\USp{4} \times \SU{2}$, respectively.
Combining this expansion with the decomposition \eqref{eq:E7Decomp} shows 
four $\USp{4} \times \SU{2}$ invariant combinations in \eqref{eq:ctY}: one at KK level $n=0$, coming from the ${[0,0,0]}$ generator, two at KK level $n=2$, coming from the noncompact generators in the ${[2,0,2]}$ and ${[0,1,0]}$, and one at KK level four from the generator in the $[0,2,0]$. Closer inspection shows that only the generators in the ${[2,0,2]}$ and ${[0,1,0]}$ belong to the $\mathfrak{sl}(8)$ subalgebra of  \eqref{eq:FR-SL8} corresponding to Freund-Rubin configurations.\footnote{It is important to note that this Freund-Rubin $\mathfrak{sl}(8)$ does not fully contain the $\mathfrak{sl}(5) \oplus \mathfrak{sl}(3)$ subgroup appearing at zero charge in \eqref{eq:E7Decomp}.}
We can thus construct a two-parameter family of $\SL{8}$ twist matrices, interpolating between the round $S^7$ and the squashed $S^7$. We choose to parametrise them as
\begin{equation}
{U}(\alpha,\eta) =
\mathring{U}(\alpha)\, e^{\eta\, T_{(5,1)}({\cal Y})} 
\;,
\label{eq:twist_squashed}
\end{equation}
with $\mathring{U}(\alpha)$ from \eqref{eq:Uround_alpha}, however with free flux parameter $\alpha$, and $T_{(5,1)}({\cal Y})$ denotes the $\USp{4} \times \SU{2}$ invariant contraction of non-compact generators in the ${[0,1,0]}$ in \eqref{eq:E7Decomp} with the round $S^7$ harmonics.
Apart from the twist matrix for the round sphere ${U}(1,0)$, the intrinsic torsion \eqref{eq:DefX} associated to \eqref{eq:twist_squashed} depends on the $S^7$ coordinates. The field equations of $D=11$ supergravity for this background can be expressed in terms of the intrinsic torsion as \eqref{eq:eom_scalars} and turn out to be identically satisfied for the values
\begin{equation}
\left\{\eta=0\,,\;\alpha=\pm1\right\}\;,\qquad
\left\{\eta=-\frac12\,{\rm log}\,5\,,\;\alpha=\pm\frac35\right\}
\;.
\label{eq:eta_alpha}
\end{equation}
It is straightforward to verify that the internal seven-dimensional metric obtained from \eqref{eq:twist_squashed} is an Einstein metric precisely for these values of the parameters. The first solution in \eqref{eq:eta_alpha} describes the round sphere and its skew-whiffed counterpart, obtained by flipping the sign of the seven-form flux. The second solution in \eqref{eq:eta_alpha} describes the left- and right-squashed spheres. With the Killing vector fields on the round sphere given by $K_{ab}{}^{m} = 2\,{\cal Y}_{[a} \partial^m {\cal Y}_{b]}$ (where the vector index on the r.h.s.\ is raised with the round $S^7$ metric), the above generalised vielbein induces the following explicit metric
\begin{equation}
g^{mn}_{(\eta)} = \frac12\,{\cal K}_{ab}{}^m {\cal K}_{ab}{}^n + \frac1{8}\,(e^{-2\eta}-1)\,  \Gamma^{ab}_{ij} \Gamma^{cd}_{ij} \, {\cal K}_{ab}{}^m {\cal K}_{cd}{}^n
\;,
\end{equation}
on the squashed sphere. Here, the $\Gamma^{ab}$ are the ${\rm SO}(8)$ $\Gamma$ matrices, with spinor indices $i, j$ running over the range $\{1, 2 , 3\},$ in accordance with the breaking \eqref{eq:USp4SU2Embedding}.

\subsection{The Kaluza-Klein spectrum on the left-squashed sphere}
We can now use the methods outlined in section~\ref{s:KKBeyondCT} to compute the full Kaluza-Klein spectrum of the squashed $S^7$. Since the squashed $S^7$ can also be described as the coset space 
\begin{equation}
\frac{\USp{4} \times \SU{2}}{\SU{2} \times \SU{2}}\;,
\label{eq:coset}
\end{equation} 
the KK spectrum can also, in principle, be computed using group theory techniques \cite{Salam:1981xd}. However, because the squashed $S^7$ is not a symmetric space, the resulting procedure is still rather intricate, although many partial results have been collected over the years~\cite{Nilsson:1983ru,Yamagishi:1983ri,Duff:1986hr,Nilsson:2018lof,Ekhammar:2021gsg,Karlsson:2021oxd}. In particular, while the set of potential mass eigenvalues of all different bosonic KK towers has been analysed to some extent (and is complete as we shall show), the traditional Kaluza-Klein computational scheme struggles to assign the eigenvalues with possible multiplicities to the correct eigenstates.

Here, we will determine the full KK spectrum on both, left- and right-squashed $S^7$ by evaluating the ExFT mass formulas. This straightforwardly provides not only the mass eigenvalues but also the corresponding eigenstates and multiplicities. In order to diagonalise the differential operators \eqref{eq:MassMatrixGraviton}, \eqref{eq:MassMatrixVector}, \eqref{eq:MassMatrixScalar}, we evaluate them on general polynomials in the fundamental round harmonics ${\cal Y}^a$. In contrast to previous ExFT computations based on \cite{Malek:2019eaz,Malek:2020yue}, the $y$-dependence of the intrinsic torsion computed from \eqref{eq:twist_squashed}, thus of the coefficients in the mass operators, induces a level mixing for the mass eigenstates. I.e.\ the action of the mass operators on a given polynomial of harmonics does not preserve the order of the polynomial. However, the symmetry group $\USp{4} \times \SU{2}$ is still large enough to keep the problem manageable. In particular, the fact that the tower of harmonics~\eqref{eq:tower_harmonics} does not carry any non-trivial multiplicities implies that any given representation appears only a finite number of times in the full KK spectrum. In order to determine its mass eigenvalues, it is thus sufficient to evaluate the corresponding mass operator on a sufficiently large polynomial of harmonics after projection onto the relevant representation. The computation can be further reduced by further projecting the polynomial onto highest weight states of $\USp{4} \times \SU{2}$. Concretely, we have pushed the computation up to Kaluza-Klein level $n=8$ which together with the underlying supersymmetry is sufficient to extract the generic structure.

In the following, we summarise our results. In appendix~\ref{app:Spectrum}, we give more details allowing us to match the results of~\cite{Nilsson:1983ru,Yamagishi:1983ri,Duff:1986hr,Nilsson:2018lof,Ekhammar:2021gsg,Karlsson:2021oxd}. We start with the left-squashed sphere, for which the states organise into long ${\cal N}=1$ supermultiplets. A generic long multiplet $L[J,\Delta]$ is identified by the (space-time) spin $J$ and the conformal dimension $\Delta$ of its superconformal primary. The conformal dimension of the primary is bounded by $\Delta > J + 1$. The long multiplets with fields of spin no higher than 2 consist of the following supergravity fields:
\begin{equation}
    \begin{split}
        L[\tfrac32,\Delta]: \psi_\mu &\xrightarrow{~Q~} g_{\mu\nu} \oplus A_\mu \xrightarrow{~Q~} \psi_\mu \,, \\
        L[1,\Delta]: A_\mu &\xrightarrow{~Q~} \,\psi_\mu\,  \oplus  \lambda\phantom{_\mu} \xrightarrow{~Q~} A_\mu \,, \\
        L[\tfrac12,\Delta]: \lambda{\phantom{_\mu}} &\xrightarrow{~Q~} \,A_\mu\,  \oplus  \phi\phantom{_\mu} \xrightarrow{~Q~} \lambda \,, \\
        L[0,\Delta]: \phi{\phantom{_\mu}} &\xrightarrow{~Q~} \phantom{A_\mu}\;\; \lambda \phantom{+\,}\;\; \xrightarrow{~Q~} \phi \,,
    \end{split}
\end{equation}
where $g_{\mu\nu}$ denotes a spin-2 field, $\psi_\mu$ denotes a spin-3/2 field, $A_\mu$ denotes a spin-1 field, $\lambda$ denotes a spin-1/2 field and $\phi$ a scalar.
In addition, the spectrum contains short multiplets $A_1[J]$ with $\Delta=1+J$, which carry the gauge fields.  These are
\begin{equation}
    \begin{split}
        A_1[\tfrac32]: \psi_\mu &\xrightarrow{~Q~} g_{\mu\nu}  \,, \\
        A_1[\tfrac12]: \lambda{\phantom{_\mu}} &\xrightarrow{~Q~} \,A_\mu\, \,.
            \end{split}
\end{equation}

The entire masse spectrum organises into a sum of long multiplets
\begin{equation}
\bigoplus L[J,\Delta] \otimes \left[p,q,r\right] \,,
\end{equation}
 in the different $\USp{4} \times \SU{2}$ representations $\left[p,q,r\right]$. Each such multiplet comes with a certain multiplicity. Remarkably, we find that the conformal dimensions of all these multiplets are captured by the universal formula
\begin{equation} \label{eq:S7Spectrum}
	\Delta_{J,s} = 1 + \frac53 s + \frac13 \sqrt{(3J+2s^2)^2 + 5\, \CasT} \,,
\end{equation}
in terms of the spin $J$  and the combination 
\begin{equation}
	\CasT = \Cas{p,q} + 3\, \Cas{r} \,,
\end{equation}
of the $\USp{4}$ and the $\SU{2}$ Casimir operators
\begin{equation}
\Cas{p,q} = \frac12 \left( p^2 + 2\, q^2 + 4\,p + 6\,q + 2\,p\,q \right)\,,\qquad
\Cas{r} = \frac14 r(r+2)
\,.
\end{equation}
The parameter $s\in\frac12\mathbb{Z}$ in \eqref{eq:S7Spectrum} is an additional label that organises the spectrum and counts the multiplicities.
To present the spectrum in compact form, we use the following notation
\begin{equation}
L[J]\smallotimes \{ s_1, s_2 ,\dots, s_p \} \equiv  \bigoplus_{i=1}^p L[J,\Delta_{J,s_i}]
\;,
\label{eq:not1}
\end{equation}
with conformal dimensions $\Delta_{J,s}$ given by \eqref{eq:S7Spectrum}. Remarkably, for all but a handful of small representations, $s$ in fact appears like the $\mathbb{R}^+ \subset \SL{2}$ charge of a full $\SL{2}$ representation. Accordingly, we use the notation
\begin{equation}
L[J]\smallotimes[S] \equiv  L[J]\smallotimes \{ -S, -S+1 ,\dots, S \}
\;.
\label{eq:not2}
\end{equation}
Let us take as an example the states in a $\left[k,q,k\right]$ of $\USp{4} \times \SU{2}$ for generic values of $k, q$ (i.e.\ $k>1$, $q>1$). The KK spectrum exhibits one spin-2 state, 9 vectors and 16 scalar fields in this representation. They turn out to fall into 13 ${\cal N}=1$ supermultiplets which in the notation \eqref{eq:not2} take the form
\begin{equation}
\left[k,q,k\right]_{k>1,q>1}  : \quad   L[\tfrac32] \smallotimes [0] 
\;\oplus\; L[1] \smallotimes [\tfrac12] 
\;\oplus\; L[\tfrac12] \smallotimes [\tfrac12\smallotimes\tfrac12] 
\;\oplus\; L[0]\smallotimes [\tfrac12\smallotimes 1]\;.
\label{eq:spectrum1}
\end{equation}
Similarly, the supermultiplets in the other generic towers of $\USp{4} \times \SU{2}$ representations 
can be summarised as
\begin{equation}
	\begin{split}
		\left[k,q,k+2\right]_{k>0,q>1} \& \left[k+2,q,k\right]_{k>0,q>0} &: \quad L[0]\smallotimes[\tfrac12]
		\; \oplus \; L[1]\smallotimes[\tfrac12]
		\; \oplus \; L[\tfrac12]\smallotimes [\tfrac12\smallotimes\tfrac12] \,, \\
		\left[k,q,k+4\right]_{q>1} \& \left[k+4,q,k\right] &: \quad L[\tfrac12]\smallotimes [0] \;\oplus\; L[0]\smallotimes [\tfrac12] \,.
	\end{split}
\label{eq:spectrum2}
\end{equation}
In appendix~\ref{app:Degeneracies}, we collect all the remaining supermultiplets in representations with small Dynkin labels, for which some of the generic structures \eqref{eq:spectrum1}, \eqref{eq:spectrum2} degenerate. Together with the universal expression \eqref{eq:S7Spectrum} for the conformal dimensions, this provides the full KK spectrum on the left-squashed sphere.

Translating the conformal dimensions into supergravity masses, we find that all mass eigenvalues fit into the list of potential eigenvalues identified in~\cite{Nilsson:1983ru,Yamagishi:1983ri,Duff:1986hr,Nilsson:2018lof,Ekhammar:2021gsg,Karlsson:2021oxd}. Seemingly missing eigenvalues on that list are explained by non-trivial multiplicities arising in the expansion of \eqref{eq:spectrum1}, \eqref{eq:spectrum2}. For vector fields, and sufficiently large values of $k$ and $q$, the masses found in \eqref{eq:spectrum1}, \eqref{eq:spectrum2} span the entire list of~\cite{Yamagishi:1983ri,Ekhammar:2021gsg}. For scalar fields, the masses realised in the KK spectrum \eqref{eq:spectrum1}, \eqref{eq:spectrum2} fix all potential sign ambiguities in the general analysis. For small values of $k$ and $q$, the general structure of the spectrum degenerates, as spelled out in \eqref{eq:degeneracies}, such that only a subset of the potential mass eigenvalues are realised. We illustrate this comparison in more detail in appendix~\ref{app:Spectrum}.

\subsection{The right-squashed sphere}

The right-squashed sphere is obtained by flipping the sign of the seven-form flux of the solution. The precise relation between the spectra on the left-squashed sphere and its ``skew-whiffed'' right-squashed counterpart can be inferred from the general results of~\cite{Duff:1983ajq,Duff:1986hr}. Combining this with the explicit form of the left-squashed spectrum \eqref{eq:spectrum1}, \eqref{eq:spectrum2}, we may describe the passage from the left-squashed to the right-squashed spectrum multiplet by multiplet (of course the right-squashed sphere breaks all supersymmetries, such that the resulting structure is no longer a supermultiplet).  In the following, we give the bosonic part of the spectrum for the right-squashed sphere. The complete result including fermions can be found in appendix D.

For all multiplets%, except those transforming as $[0,0,0]$, $[1,0,1]$ and $[0,1,0]$
, we find the following picture. First of all, the bosonic masses in all spin-2 multiplets as well as in vector multiplets, i.e.\ $L[\tfrac32]$ and $L[1]$ multiplets, remain unchanged. 
For the bosonic states of the $L[\tfrac12]$ multiplets, the transition works as follows 
\begin{equation}
	\begin{split}
		L[\tfrac12]\smallotimes \{-1 \} &: \quad \left\{  \begin{array}{l}
			\mathrm{vector} : \Delta_{\rm RS} = \Delta_{\rm LS} \,,\\[0.5em]
			\mathrm{scalar} : \Delta_{\rm RS} = \Delta_{\rm LS} \EM{~+~ 2|s|+1} \,,
		\end{array} \right. \\
		L[\tfrac12]\smallotimes \{1 \} &: \quad \left\{  \begin{array}{l}
			\mathrm{vector} : \Delta_{\rm RS} = \Delta_{\rm LS} \,, \\[0.5em]
			\mathrm{scalar} : \Delta_{\rm RS} = \Delta_{\rm LS} \EM{~-~ 2|s|-1} \,,
		\end{array} \right. \\	\end{split}
\end{equation}
while the bosonic states of $L[\tfrac12]\smallotimes \{0 \} $ remain unchanged. This is illustrated in Fig. \ref{fig:LeftRight12bos}.
\begin{figure}[h!]
\begin{center}
\begin{tikzpicture}
        % Titres des zones
    \node at (0,10) {$L[\tfrac12] \smallotimes \{-1 \}$ };
    \node at (5,10) {$ L[\tfrac12]\smallotimes \{ 1\}$};
    
    \node at (-1,9) {LS};
    \node at (1,9) {RS};
    \node at (4,9) {LS};
    \node at (6,9) {RS};
    \node at (-1,7.4) {$\left\{A_{\mu},\phi\right\}$};
    \node at (1,7.4) {$A_{\mu}$};
    \node at (1,5.6) {$\phi$};
    \node at (4,6.6) {$\left\{A_{\mu},\phi\right\}$};
    \node at (6,8.4) {$\phi$};
    \node at (6,6.6) {$A_{\mu}$};
    
    \node at (6.5,7.5) {\EM{+3}};
    \node (ZA) at (1.5,6.5) {\EM{-3}};
    
    \draw (-2,9.5) -- (7,9.5);
    \draw (2.5,5) -- (2.5,10.5);
    
    % Lignes 
    \draw [->] (-1,7) -- (1,7) ;
    \draw [->] (-1,7) -- (1,6) ;
    \draw [->] (4,7) -- (6,8) ;
    \draw [->] (4,7) -- (6,7) ;

  % Lignes 
    \draw[red,very thick] [->] (1.1,6.9) -- (1.1,6.1) ;
    \draw[red,very thick] [->] (6.1,7.1) -- (6.1,7.9) ;
    
    % Points
    \draw (-1,7) node{$\bullet$} ;
    \draw (1.1,6) node{$\bullet$} ;
    \draw (1.1,7) node{$\bullet$} ;
    \draw (4,7) node{$\bullet$} ;
    \draw (6.1,7) node{$\bullet$} ;
    \draw (6.1,8) node{$\bullet$} ;
\end{tikzpicture}
\end{center}
\caption{Shift patterns of the conformal dimensions of the bosonic states within $L[\frac12]$, as we go from the left- to the right-squashed $S^7$.} \label{fig:LeftRight12bos}
\end{figure}
Between the left and right squashing, the scalar is shifted by $\pm 3$ in the direction of ${\rm sign}(-s)$\,. As a result, in the case of the right squashing, the scalar and vector states which used to have the same conformal dimension are now separated by $\pm 3$ in the right-squashed sphere, again depending on the sign of $s$. 

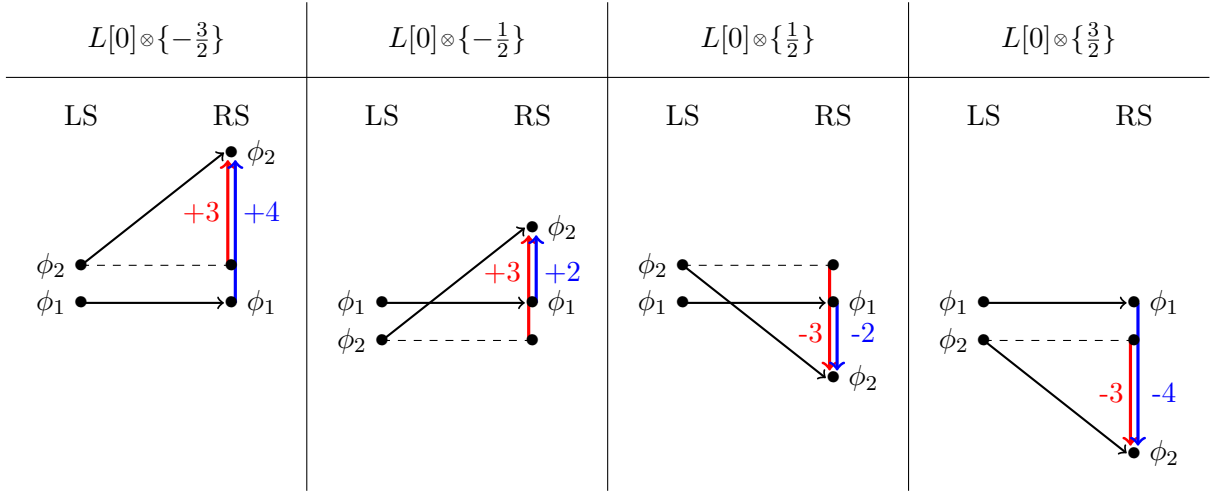
\begin{figure}[h!]
\begin{center}
\begin{tikzpicture}
        % Titres des zones
    \node at (0,10) {$L[0] \smallotimes \{-\tfrac32 \}$};
    \node at (4,10) {$L[0] \smallotimes \{-\tfrac12 \}$};
    \node at (8,10) {$L[0] \smallotimes \{\tfrac12 \}$};
    \node at (12,10) {$L[0] \smallotimes \{\tfrac32 \}$};
    
    \node at (-1,9) {LS};
    \node at (1,9) {RS};
    \node at (-1.4,6.5) {$\phi_1$};
    \node at (-1.4,7) {$\phi_2$};
    \node at (1.4,6.5) {$\phi_1$};
    \node at (1.4,8.5) {$\phi_2$};
    \node at (3,9) {LS};
    \node at (5,9) {RS};
    \node at (2.6,6.5) {$\phi_1$};
    \node at (2.6,6) {$\phi_2$};
    \node at (5.4,6.5) {$\phi_1$};
    \node at (5.4,7.5) {$\phi_2$};
    \node at (7,9) {LS};
    \node at (9,9) {RS};
    \node at (6.6,6.5) {$\phi_1$};
    \node at (6.6,7) {$\phi_2$};
    \node at (9.4,6.5) {$\phi_1$};
    \node at (9.4,5.5) {$\phi_2$};
    \node at (11,9) {LS};
    \node at (13,9) {RS};
    \node at (10.6,6.5) {$\phi_1$};
    \node at (10.6,6) {$\phi_2$};
    \node at (13.4,6.5) {$\phi_1$};
    \node at (13.4,4.5) {$\phi_2$};

    \node at (0.6,7.7) {\EM{+3}};    
    \node at (1.4,7.7) {\BL{+4}};    
    \node at (4.6,6.9) {\EM{+3}};
    \node at (5.4,6.9) {\BL{+2}};
    \node at (8.7,6.1) {\EM{-3}};
    \node at (9.4,6.1) {\BL{-2}};
    \node at (12.7,5.3) {\EM{-3}};
    \node at (13.4,5.3) {\BL{-4}};
    
    \draw (2,4) -- (2,10.5);
    \draw (6,4) -- (6,10.5);
    \draw (10,4) -- (10,10.5);
    \draw (-2,9.5) -- (14,9.5);
    
    % Lignes 
    \draw[thick] [->] (-1,6.5) -- (0.9,6.5) ;
    \draw[thick] [->] (-1,7) -- (0.9,8.5) ;
    \draw[dashed] (-1,7) -- (1,7) ;
    \draw[thick] [->] (3,6.5) -- (4.9,6.5) ;
    \draw[thick] [->] (3,6) -- (4.9,7.5) ;
    \draw[dashed] (3,6) -- (5,6) ;
    \draw[thick] [->] (7,6.5) -- (8.9,6.5) ;
    \draw[thick] [->] (7,7) -- (8.9,5.5) ;
    \draw[dashed] (7,7) -- (9,7) ;
    \draw[thick] [->] (11,6.5) -- (12.9,6.5) ;
    \draw[thick] [->] (11,6) -- (12.9,4.5) ;
    \draw[dashed] (11,6) -- (13,6) ;
  % Lignes 
    \draw[red,very thick] [->] (0.95,7) -- (0.95,8.4) ;
    \draw[blue,very thick] [->] (1.05,6.5) -- (1.05,8.4) ;
    \draw[red,very thick] [->] (4.95,6) -- (4.95,7.4) ;
    \draw[blue,very thick] [->] (5.05,6.5) -- (5.05,7.4) ;
    \draw[red,very thick] [->] (8.95,7) -- (8.95,5.6) ;
    \draw[blue,very thick] [->] (9.05,6.5) -- (9.05,5.6) ;    
    \draw[red,very thick] [->] (12.95,6) -- (12.95,4.6) ; 
    \draw[blue,very thick] [->] (13.05,6.5) -- (13.05,4.6) ;        
    % Points
    \draw (-1,6.5) node{$\bullet$} ;
    \draw (-1,7) node{$\bullet$} ;
    \draw (1,7) node{$\bullet$} ;
    \draw (1,6.5) node{$\bullet$} ;
    \draw (1,8.5) node{$\bullet$} ;
   
    \draw (3,6.5) node{$\bullet$} ;
    \draw (3,6) node{$\bullet$} ;
    \draw (5,7.5) node{$\bullet$} ;
    \draw (5,6.5) node{$\bullet$} ;
    \draw (5,6) node{$\bullet$} ;
    
    \draw (7,6.5) node{$\bullet$} ;
    \draw (7,7) node{$\bullet$} ;
    \draw (9,5.5) node{$\bullet$} ;
    \draw (9,6.5) node{$\bullet$} ;    
    \draw (9,7) node{$\bullet$} ;
     
    \draw (11,6) node{$\bullet$} ;
    \draw (11,6.5) node{$\bullet$} ;
    \draw (13,6.5) node{$\bullet$} ;
    \draw (13,4.5) node{$\bullet$} ;  
    \draw (13,6) node{$\bullet$} ;
      
\end{tikzpicture}
\end{center}
\caption{Shift patterns of the conformal dimensions of the bosonic states within $L[0]$, as we go from the left- to the right-squashed $S^7$.} \label{fig:LeftRight0bos}
\end{figure}

The $L[0]$ multiplets behave similarly. First, there is always a scalar state, whose conformal dimension is unchanged between the left and the right squashing. The other scalar state gets shifted by $\pm 3$ depending on the sign of $s$. In order to identify which of the states gets shifted, one notes that for the right squashing, the difference in the conformal dimension between bosonic states is no longer 1 but changes to $2|s|+1$ as for the $L[\frac12]$ multiplets. This can be summed up as in Fig. \ref{fig:LeftRight0bos}.
Again, the conformal dimensions of half of the states are shifted by $\pm 3$ between the right and left squashing, such that the conformal dimensions of the bosonic states from the same multiplet now differ by an $s$ dependent shift.

\subsection{Rational conformal dimensions and marginal deformations}
\begin{figure}[tb]
    \centering
    \includegraphics[scale=0.2]{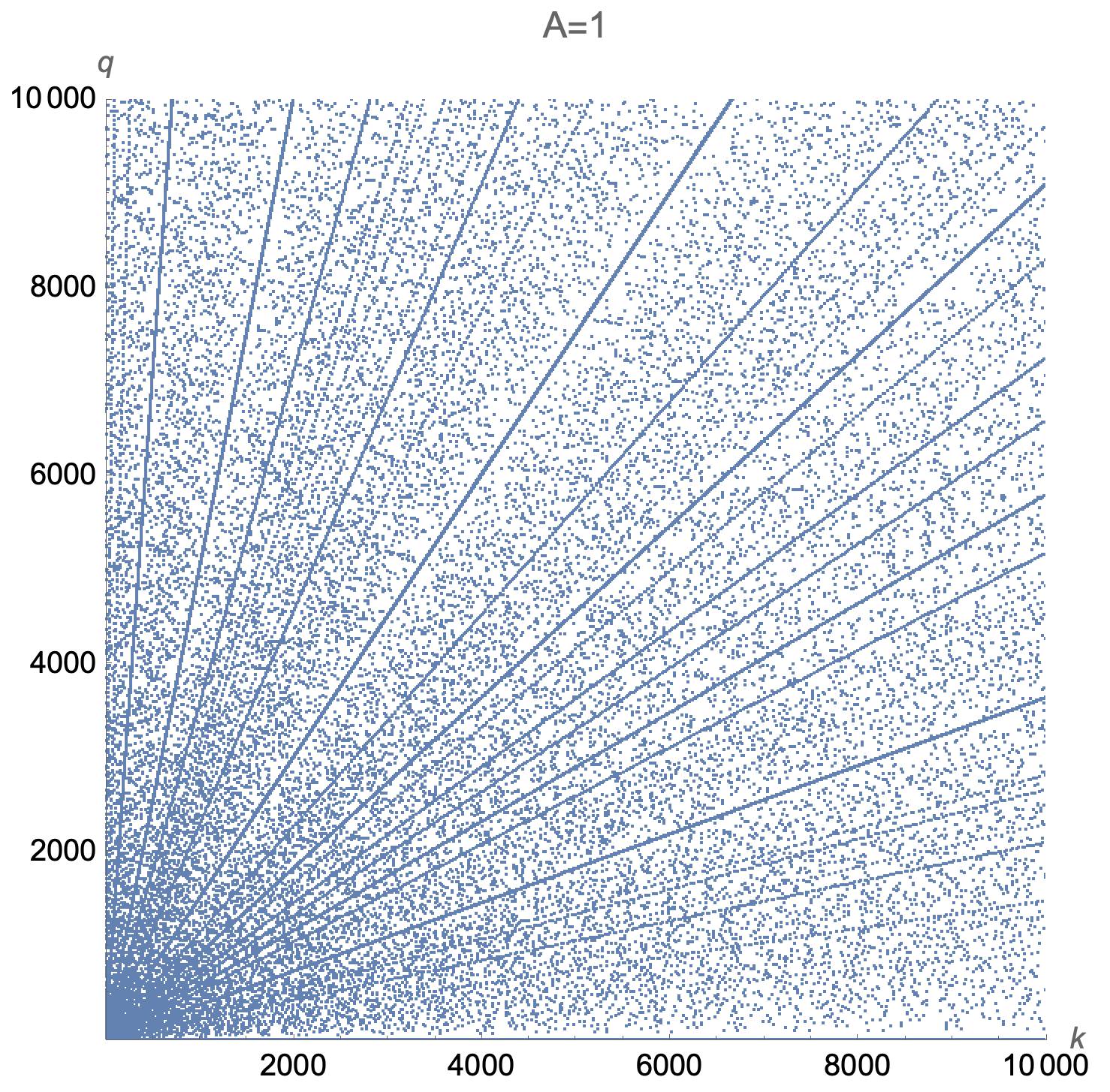}
    \includegraphics[scale=0.2]{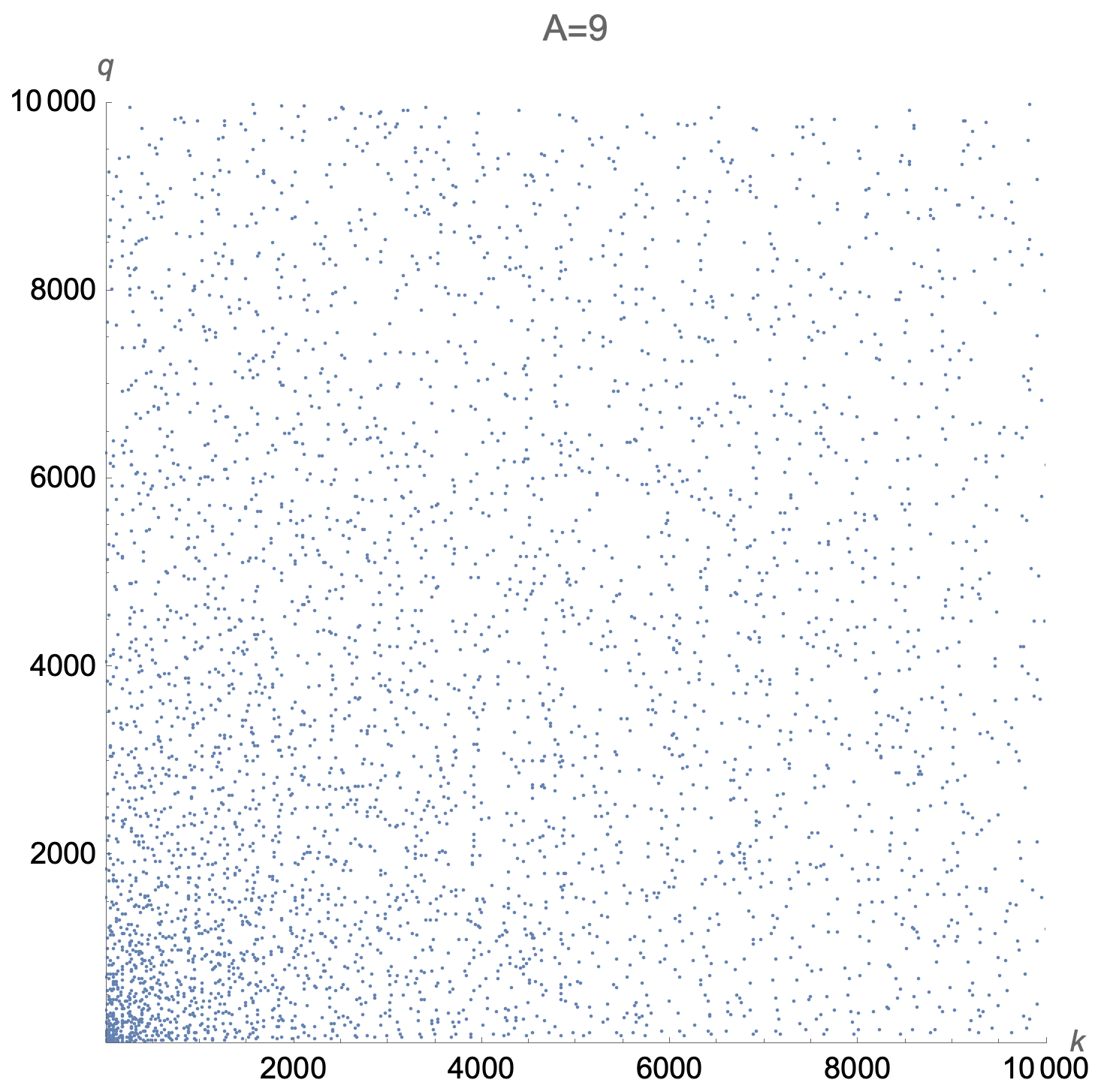}
    \includegraphics[scale=0.2]{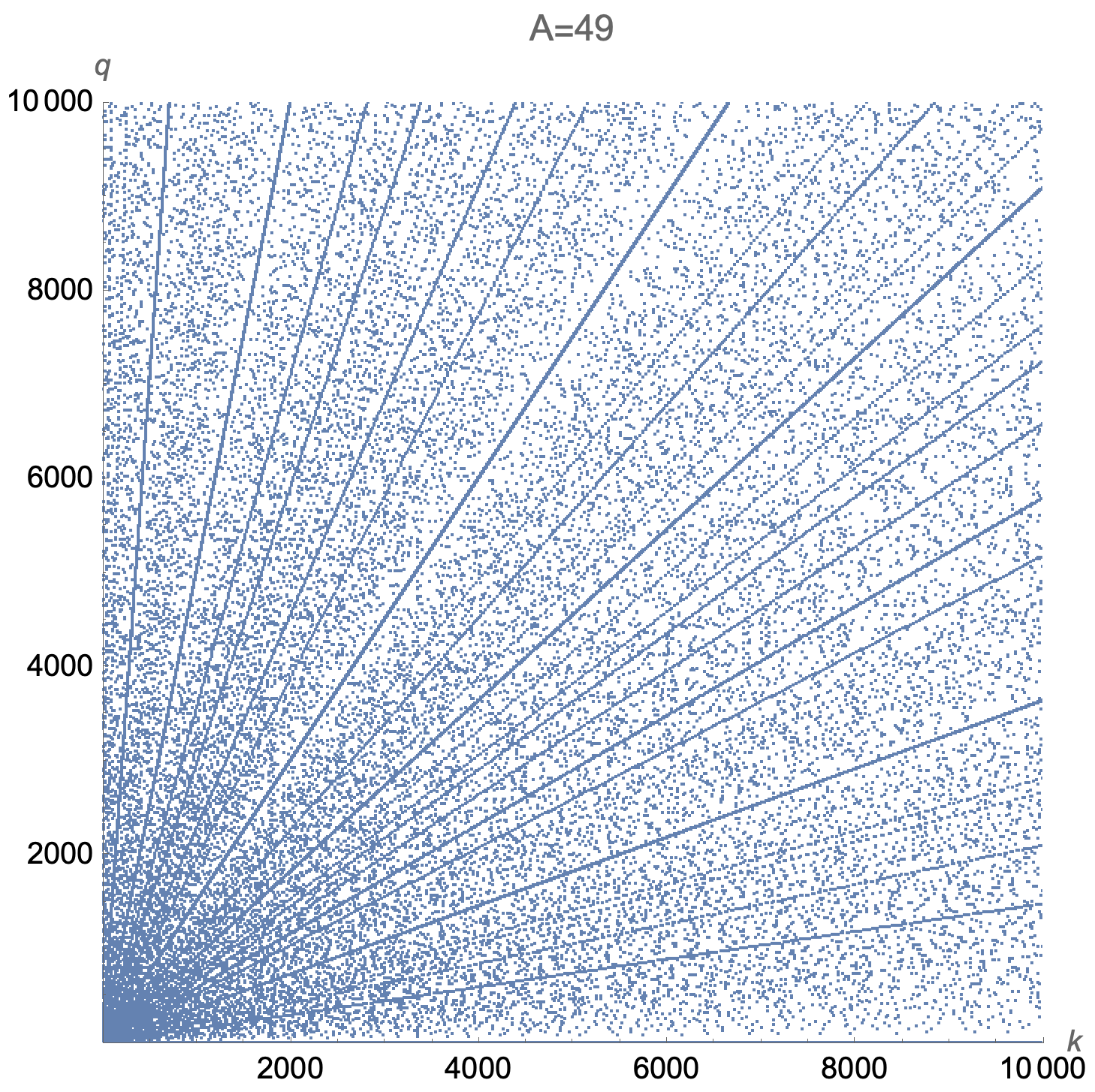}
    \includegraphics[scale=0.2]{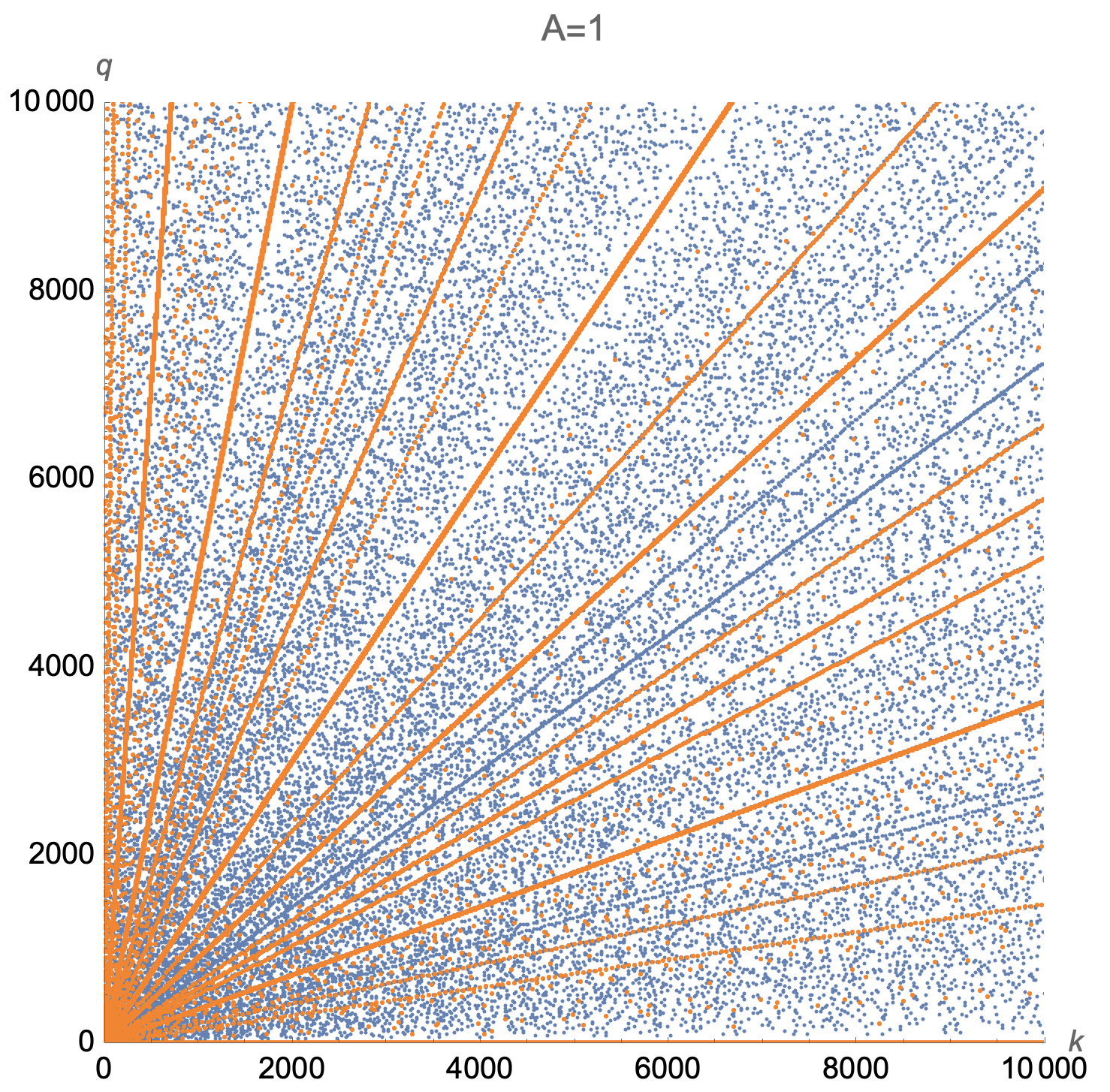}
    \includegraphics[scale=0.2]{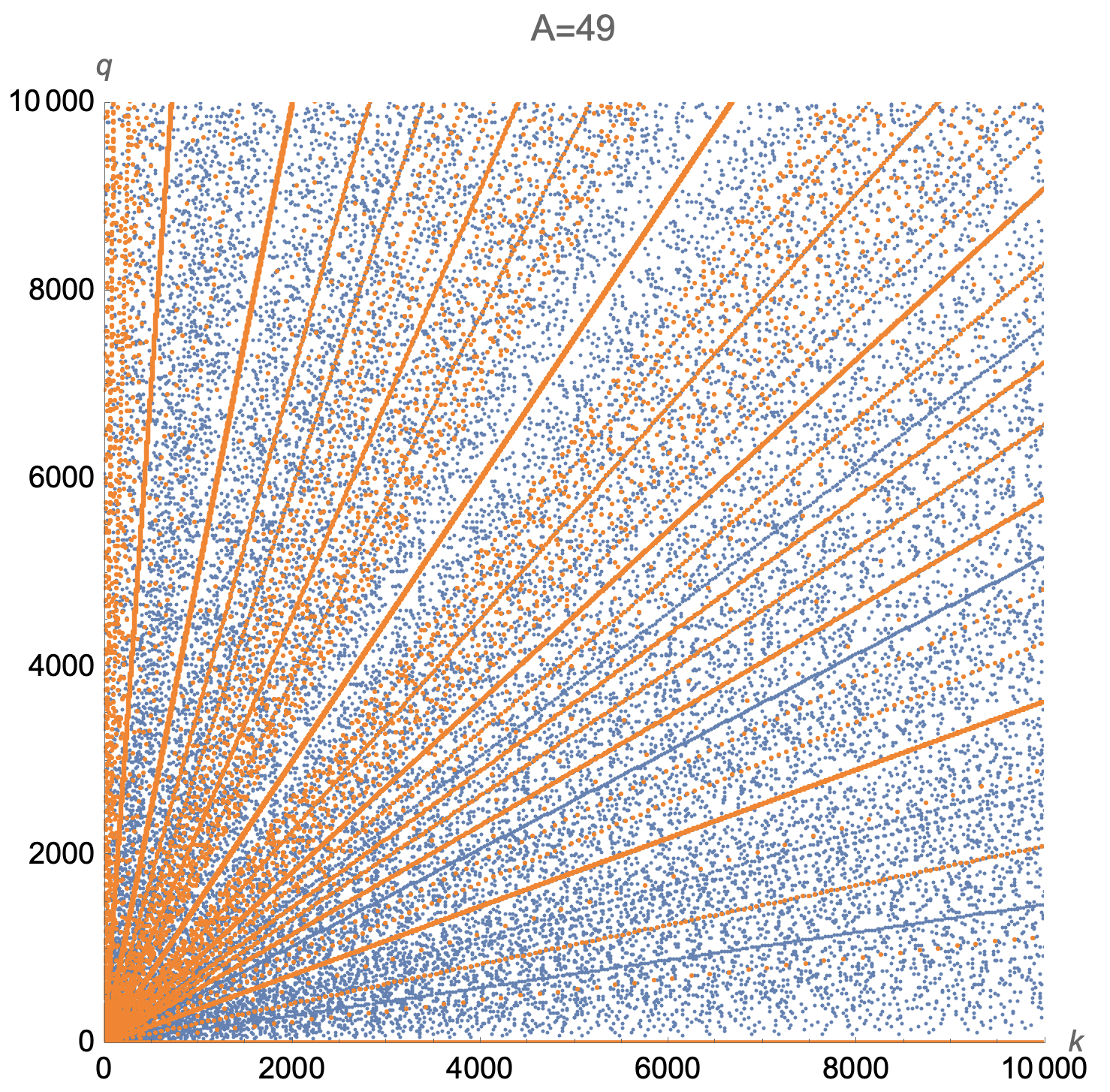}
        \caption{In the first row of figures, the blue points give integer solutions to \eqref{eq:MustBeRational} for the special values $A=1,9,49$. In the second row, 
        we superpose these plots with orange lines corresponding to analytical solutions to the Pell equation  \eqref{eq:PellEquation} as discussed in the text.}
    \label{fig:kqplot}
\end{figure}

Since the AdS$_4$ vacuum only preserves ${\cal N}=1$ supersymmetry, all multiplets are unprotected, i.e.\ they are either long or sit at the unitarity bound where they can recombine into long multiplets. Still, we observe infinitely many rational conformal dimensions in the KK spectrum. In particular, these arise from the following towers (in the notation of \eqref{eq:not1})
\begin{equation}\label{eq:RatConfDim}
	\begin{split}
		L[0] \smallotimes \{ -\tfrac12, \tfrac12 \}  \otimes \left[k,1,k\right]_{k>1} &: \quad \left\{ \begin{array}{l}
			\Delta = \frac{10+5k}{6} \,, \\[0.5em]
			\Delta = \frac{20+5k}{6} \,,
		\end{array}  \right. \\
		L[1]  \smallotimes \{ - \tfrac12 \}   \otimes \left[k,0,k\right]_{k>1}&: \quad \Delta = \frac{8+5k}{6} \,, \\
		L[\tfrac12] \smallotimes \{ 1 \}   \otimes \left[k,0,k\right]_{k>1} &: \quad \Delta = \frac{23+5k}{6} \,, \\
		L[1] \smallotimes  \{ \tfrac12 \} \otimes \left[k,0,k+2\right]  &: \quad \Delta = \frac{24+5k}{6}\,, \\
		L[\tfrac12]\smallotimes  \{ -1 \}   \otimes \left[k,0,k+2\right]  &: \quad \Delta = \frac{9+5k}{6}\,, \\
		L[0]  \smallotimes  \{ \tfrac12 \}  \otimes \left[k+2,0,k\right] &: \quad \Delta = \frac{22+5k}{6}\,, \\
		L[0] \smallotimes \{ -\tfrac12 \} \otimes \left[k,1,k+4\right]  &: \quad \Delta = \frac{22+5k}{6}\,, \\
		L[0]  \smallotimes \{ -\tfrac12 \}  \otimes \left[k,0,k+4\right] &: \quad \Delta = \frac{20+5k}{6}\,,
	\end{split}
\end{equation}
whose conformal dimensions are manifestly rational.
More general, we can study rational solutions of \eqref{eq:S7Spectrum}. In order for the conformal dimension to be rational, this requires 
\begin{equation} \label{eq:MustBeRational}
\sqrt{A + 70 k + 25 k ^2 + 60 q + 20 k q + 20 q^2} ~\in~ \mathbb{N}\;,
\end{equation}
with $A \in \{1,9,49,81\}$. In other words, we need to solve the following order two diophantine equation\footnote{
Similar structures have been revealed by Gubser \cite{Gubser:1998vd} in the KK spectrum on type IIB supergravity on AdS$_5 \times T^{1,1}$, see also \cite{Ceresole:1999zs}.}
\begin{equation} \label{eq:Diophantine}
A + 70 k + 25 k ^2 + 60 q + 20 k q + 20 q^2-N^2=0\;.
\end{equation}
In order to get a feeling on general rational solutions of \eqref{eq:MustBeRational}, we may numerically plot integer solutions of \eqref{eq:MustBeRational} in the $(k,q)$ plane. The results are shown in Fig. \ref{fig:kqplot}. We can see for the values $A=1$ and $A=49$ lines emerging from the graph, whereas there are no such lines on the graph $A=9$ (and similarly for $A=81$). As a consequence, we look for solutions of the form $q=a k + b, (a,b) \in \mathbb{Q}^2$. Plugging this into \eqref{eq:MustBeRational} we find an order two polynomial in $k$, whose discriminant $\Delta$ must vanish. As $\Delta$ is a function of $a$ and $b$, we can solve the equation $\Delta=0$ for $a$. In order to have $a \in \mathbb{Q}$ we must find $b$ such that $\sqrt{-(A-65) (A + 20 b (3 + b))} \in \mathbb{Q}$. For $A=81$, the number in the square root is always a negative number, which explains why we do not see any line in the $A=81$ plot. For $A=1$ and $A=49$, the first factor $(A-65)$ gives an exact square number and we must find $b$ such that $\sqrt{A + 20 b (3 + b)} \in \mathbb{Q}$. The problem finally reduces to finding $b$ such that $A + 60 b + 20 b^2=y^2$, $y \in \mathbb{Q}$, and substituting $x=b-\frac32$, we must solve 
\begin{equation}\label{eq:PellEquation}
A -45  + 20 x^2 - y^2 = 0\;.
\end{equation}
This is a Pell equation, whose integers solutions can be found using %$\mathrm{Reduce}[4 + 20 x^2 - y^2 == 0 \&\& \mathrm{Element}[x | y \mathrm{Integers}]]$ in 
Mathematica. However, we are not only interested in integer solutions, but also in rational solutions of this equation. We must solve $y^2-20\,x^2=A-45$. In order to find solutions, we first solve what we will call the homogeneous Pell equation $y^2-D\,x^2=1$. It can be shown that rational solutions of the homogeneous Pell equation can be written as $(x,y)=(\frac{t^2+1}{t^2-1},\, \frac{2t}{t^2-1}),\, t \in \mathbb{Q},\, t^2\ne D$. Solutions to the original Pell equation can eventually be found using a particular solution, and multiplying it by the homogeneous solutions. Indeed, let $(x_0,y_0)$ be a particular solution and $(x,y)$ a solution to the homogeneous Pell equation, then $A-45=(y_0^2-Dx_0^2)(y^2-Dx^2)=(x_0x\pm y_0y)^2-D(x_0y\pm y_0x)^2$, allowing us to generate families of solutions of the Pell equation. This method works as long as $D$ is not a square number. We also want to emphasise that this may not be all solutions of the Pell equation, as a different particular solution may lead to a different family of solutions. We illustrate our findings with the orange lines in Fig. \ref{fig:kqplot} for the case $A=1$ and $A=49$.

Particularly interesting are the multiplets with marginal deformations. For the ${\cal N}=1$ left-squashed sphere these are
\begin{equation}
		L[0]\smallotimes \{-\tfrac32\} \otimes[0,3,0]\,, 
		\quad L[0]\smallotimes \{-\tfrac32 \}\otimes[2,1,2] \,.
		\label{eq:marginalLS}
\end{equation}
These are $D$-terms and preserve the ${\cal N}=1$ supersymmetry. We note that one of the massless scalars preserves the $\USp{4} \times \SU{2}$ symmetry, while the second one breaks $\USp{4}$ to $\SU{2} \times \SU{2}$ and preserves $\SU{2}$.
For the right-squashed sphere, all massless scalars in \eqref{eq:marginalLS} turn massive. However, massive scalars from the following multiplets of the left-squashed sphere
\begin{equation}
\begin{split}
		L[\tfrac12]\smallotimes \{1 \} \otimes[2,1,0]\,, \quad L[\tfrac12]\smallotimes \{1 \} \otimes[2,0,2]\,,
\end{split}
\end{equation}
become massless for the right-squashed sphere by the pattern displayed in Fig. \ref{fig:LeftRight12} above. It would be very interesting to study whether any of these massless scalars for the left-/right-squashed sphere can be integrated up to finite moduli.

\section{Conclusions} \label{s:Conclusions}
In this paper, we showed how to compute the full Kaluza-Klein spectrum of supergravity compactifications which are not part of a consistent truncation, but are still generalised parallelisable. Examples of such vacua are deformations of compactifications within ${\cal N}=8$ supergravity by scalar fields which are not part of the ${\cal N}=8$ truncations. This includes the supergravity duals of RG flows of ${\cal N}=4$ SYM or ${\cal N}=8$ ABJM triggered by single-trace operators. Thus, our formalism can be used to compute the Kaluza-Klein spectrum for the end-point of such flows or even along such flows.

As an application of our method, we computed the full spectrum of the AdS$_4 \times$ squashed $S^7$ solution of 11-dimensional supergravity. This preserves only ${\cal N}=1$ supersymmetry (or ${\cal N}=0$ in the case of the right-squashed $S^7$), and thus has no protected operators. This background is a coset space, so that traditional techniques \cite{Salam:1981xd} can in principle be used. However, because the squashed $S^7$ is not a symmetric space, this is still rather difficult and had not been completed until now, despite however many explicit results~\cite{Nilsson:1983ru,Yamagishi:1983ri,Duff:1986hr,Nilsson:2018lof,Ekhammar:2021gsg}. Using our technology, we were able to straightforwardly compute the full Kaluza-Klein spectrum, which is captured by the remarkably simple formula \eqref{eq:S7Spectrum}, which depends only on the $\USp{4} \times \SU{2}$ representation of the multiplet, the spin of the superconformal primary and the charge under an additional $\mathbb{R}^+$ factor. Intriguingly, for all but the smallest representations, this $\mathbb{R}^+$ charge appears to descend from representations of a bonus $\SL{2}$, whose origin is mysterious.

The method we describe here opens up the possibility of computing the Kaluza-Klein spectrum of many more interesting string compactifications which do not reside in a ${\cal N}=8$ consistent truncation. This includes TsT transformations of vacua in ${\cal N}=8$ supergravity, such as the marginal deformations of AdS$_5 \times S^5$, AdS$_4 \times S^7$ or ${\cal N}=1$ AdS$_4$ vacua of IIB string theory \cite{Bobev:2021rtg}. Our method might also apply to the cubic deformation of AdS$_5 \times S^5$, which has recently been described implicitly in generalised geometry \cite{Ashmore:2021mao}. It would be interesting to see if this implicit description may still be sufficient to apply our method here.

Finally, it would be interesting to better understand our results from a CFT perspective. It is remarkable that the spectrum of the squashed $S^7$, which only preserves ${\cal N}=1$ supersymmetry and thus has no protected multiplets, displays such a simple structure.  A particularly interesting questions is the origin of the additional $s$ charge (and its enhancement to a bonus $\SL{2}$ group) which appears to organise the spectrum. Perhaps computing the spectrum along the RG flow from the round $S^7$ to the squashed one may shed some light into this.

\section{Acknowledgements}
We are grateful to Davide Cassani, Michele Galli, Joel Karlsson, Bengt Nilsson, Michela Petrini, Ergin Sezgin and Daniel Waldram for useful discussions and correspondence. EM is supported by the Deutsche Forschungsgemeinschaft (DFG, German Research Foundation) via the Emmy Noether program ``Exploring the landscape of string theory flux vacua using exceptional field theory'' (project number 426510644).

\appendix

\section{Mass operators for $\En{6}$ ExFT} \label{app:E6}
Here we also give the mass operators applicable to generalised parallelisable (but not necessarily Leibniz) compactifications to five dimensions. We do this using the $\En{6}$ ExFT~\cite{Hohm:2013vpa}, which has the bosonic field content
\begin{equation} \label{eq:ExFTfieldsE6}
	\left\{ g_{\mu\nu},\, \gM_{MN},\, {\cal A}_{\mu}{}^M,\, {\cal B}_{\mu\nu\,M} \right\} \,,\qquad \mu, \nu = 0, \dots, 4\,, \quad M = 1, \dots, 27 \,,
\end{equation}
where $g_{\mu\nu}$ is now the five-dimensional metric, the generalised metric $\gM_{MN}$ parameterises the coset space $\En{6}/\USp{8}$ and the indices $M, N = 1, \ldots 27$ label the fundamental of $\En{6}$. Just as in the $\En{7}$ case we consider a generalised parallelisable background, i.e. admitting a globally well-defined $\En{6}$ twist matrix $U_{\fl{A}}{}^M$ and nowhere-vanishing scalar density $\rho^{-1}$. In terms of these, we have the fluctuation Ansatz
\begin{equation}
\begin{split} \label{eq:KKAnsatzE6}
	g_{\mu\nu}(x,Y) &= \rho^{-2}(y) \Big( \vg_{\mu\nu}(x) + \sum_\Sigma \cY_\Sigma\, h_{\mu\nu}{}^{\Sigma}(x) \Big) \,, \\
	\cA_\mu{}^M(x,Y) &= \rho^{-1}(y)\, \UI_\fl{A}{}^M(y) \sum_\Sigma \cY_\Sigma(y)\, A_\mu{}^{\fl{A},\Sigma}(x) \,, \\
	\cB_{\mu\nu\,M}(x,Y) &= \rho^{-2}(y)\, U_M{}^\fl{A}(y) \sum_\Sigma \cY_\Sigma(y)\, B_{\mu\nu\,\fl{A}}{}^{\Sigma}(x) \,, \\
	\gM_{MN}(x,Y) &= U_M{}^\fl{A}(y)\, U_N{}^\fl{B}(y) \Big( \delta_{\fl{A}\fl{B}} + \cP_{I,\fl{A}}{}^{\fl{B}} \sum_\Sigma \cY_\Sigma(y)\, j^{I,\Sigma}(x) \Big) \,,
\end{split}
\end{equation}
with $\cP_{I,\fl{A}}{}^{\fl{B}}$ corresponding to the non-compact generators of $\En{6}$, with $I = 1, \ldots, 42$, which we raise and lower with the non-compact part of the $\En{6}$ Cartan-Killing metric.

The mass operators can be computed completely analogously to the $\En{7}$ case and we give them here without re-iterating the derivation. The spin-2 mass operator takes exactly the same form as \eqref{eq:MassMatrixGraviton}
\begin{equation} \label{eq:MassMatrixGravitonE6}
	\MGrav = - \partial_{\fl{A}} \partial_{\fl{A}} \,,
\end{equation}
acting on scalar harmonics $\cY_{\Sigma}$. The vector mass matrix is given by
\begin{equation} \label{eq:MassMatrixVectorE6}
	\MVec{}^{\fl{A}}{}_{\fl{B}} = \frac{1}{24}\, \Pi_{\fl{A},I}\, \Pi^I{}_{\fl{B}} \,,
\end{equation}
with
\begin{equation}
	\Pi^I{}_{\fl{A}} = - 2 \left( X_{\fl{A}}{}^I - 6\, \cP^I{}_{\fl{A}}{}^{\fl{B}}\, \partial_{\fl{B}} \right) \,,
\end{equation}
and its adjoint
\begin{equation}
	\Pi_{\fl{A},I} = - 2 \left( X_{\fl{A}I} + 6\, \cP_{I,\fl{A}}{}^{\fl{B}}\, \partial_{\fl{B}} \right) \,.
\end{equation}
The scalar mass matrix is given by
\begin{equation} \label{eq:MassMatrixScalarE6}
	\begin{split}
		\MScal{}^I{}_J &= X_{\uA\uE}{}^{\uF} X_{\uB\uF}{}^{\uE} \, (\cP^I\cP_J)_{\uA}{}^{\uB} \\
		& \quad + \frac15 \left( X_{\uA\uE}{}^{\uF} X_{\uB\uE}{}^{\uF} + X_{\uE\uA}{}^{\uF} X_{\uE\uB}{}^{\uF} + X_{\uE\uF}{}^{\uA} X_{\uE\uF}{}^{\uB}  \right) (\cP^I\cP_J)_{\uA}{}^{\uB} \\
		& \quad + \frac25 \left( X_{\uA\uC}{}^{\uE} X_{\uB\uD}{}^{\uE} - X_{\uA\uE}{}^{\uC} X_{\uB\uE}{}^{\uD} - X_{\uE\uA}{}^{\uC} X_{\uE\uB}{}^{\uD} \right) (\cP^I)_{\uA}{}^{\uB}\, (\cP_J)_{\uC}{}^{\uD} \\
		& \quad - 2\, (\cP_J)_{C}{}^{D} \, \partial_C X_{\fl{D}}{}^I -\big[ \cP^I,\cP_J\big]{}_{\fl{A}}{}^{\fl{B}}\, \partial_{\fl{C}}X_{\fl{CB}}{}^{\fl{A}} \\
		&\quad + 2\, \Big( (\cP^I)_{\fl{A}}{}^{\fl{B}} X_{\fl{A}J} - (\cP_J)_{\fl{A}}{}^{\fl{B}} X_{\fl{A}}{}^{I} \Big) \, \partial_B - 2\, \big[ \cP^I,\cP_J\big]{}_{\fl{A}}{}^{\fl{B}} 
		\,X_{\fl{CB}}{}^{\fl{A}} \, \partial_{\fl{C}} \\
		& \quad -\delta^I_J\,\partial_{\fl{A}} \partial_{\fl{A}} + 12\, (\cP^I\cP_J)_{\fl{A}}{}^{\fl{B}}\, \partial_{\fl{B}} \partial_{\fl{A}} \,.
	\end{split}
\end{equation}
Finally, the mass matrix for the 2-form can be computed from its first-order equation of motion, to obtain
\begin{equation}
	\MTForm{}_{\fl{A}}{}^{\fl{B}} = \frac{1}{\sqrt{10}} \left( - Z^{\fl{A}\fl{B}} + 10\, d^{\fl{A}\fl{B}\fl{C}}\, \partial_{\fl{C}} \right) \,.
\end{equation}
Here, $d^{\fl{A}\fl{B}\fl{C}}$ is the symmetric cubic invariant of $\En{6}$, normalised as in \cite{Malek:2020yue}, and we used the antisymmetric combination of the intrinsic torsion given by
\begin{equation}
	Z^{\fl{A}\fl{B}} = 2\, d^{\fl{C}\fl{D}\fl{A}}\, X_{\fl{C}\fl{D}}{}^{\fl{B}} = - Z^{\fl{B}\fl{A}} \,.
\end{equation}

\section{Details of the squashed $S^7$ spectrum} \label{app:Spectrum}

It is instructive to compare the above results for the KK spectrum on the squashed spheres to the results obtained in the traditional computational scheme~\cite{Nilsson:1983ru,Yamagishi:1983ri,Duff:1986hr,Nilsson:2018lof,Ekhammar:2021gsg,Karlsson:2021oxd}. In that approach, the eigenvalue spectra of the different internal Laplacian operators on the squashed sphere are determined by the coset space techniques based on the representation \eqref{eq:coset} of the internal space. These spectra are then combined with the universal formulas for mass operators appearing in Freund-Rubin compactifications~\cite{Duff:1986hr}. While extensive knowledge of the Laplacian eigenvalue spectra has been accumulated in~\cite{Nilsson:1983ru,Yamagishi:1983ri,Duff:1986hr,Nilsson:2018lof,Ekhammar:2021gsg,Karlsson:2021oxd}, the assignment of these eigenvalues and their multiplicities to specific mass eigenstates appears less straightforward in that approach.

Let us consider as an example the KK states in the $\left[k,q,k\right]$ representation (for generic values $k>1, q>1$) for the ${\cal N}=1$ left-squashed sphere. From \eqref{eq:spectrum1}, we find the multiplet structure
\begin{equation}
\left[k,q,k\right]_{k>1,q>1}  \;\;:\;\;   L[\tfrac32] \smallotimes [0] 
\;\oplus\; L[\tfrac32] \smallotimes [\tfrac12] 
\;\oplus\; L[\tfrac12] \smallotimes [\tfrac12\smallotimes\tfrac12] 
\;\oplus\; L[0]\smallotimes [\tfrac12\smallotimes 1]\;.
\label{eq:spectrum1-app}
\end{equation}
Formula \eqref{eq:S7Spectrum} yields the conformal dimensions of all fields. We may translate them into supergravity masses by the standard $D=4$ formulas
\begin{equation}
\begin{split}
\mbox{spin-0, 2} : &\quad  \Delta(\Delta-3) = m^2\,\ell^2 \;, \\
\mbox{spin-1} : & \quad (\Delta-1)(\Delta-2) = m^2\,\ell^2 \;.
\end{split}
\end{equation}
In our conventions, and with the twist matrix from \eqref{eq:twist_squashed} the AdS length $\ell$ for the squashed $S^7$ is given by $\ell^2=\frac5{72}$\,.
Evaluating the field content of the various supermultiplets in \eqref{eq:spectrum1-app}, we obtain the following masses for the different spin-2 and spin-1 modes
\begin{equation}
\begin{tabular}{c|cccc}
 & $\Delta$ & $m^2$ & $L[J]$  & \#  \\ \hline\hline
 $g_{\mu\nu}$ & $\frac32 + \frac16 \sqrt{81 + 20\, {\cal C}_3}$  & $8\,{\cal C}_3$ & $L[\frac32]$ & 1\\ \hline\hline
 $A_\mu$ & $\frac32 + \frac16 \sqrt{81 + 20\, {\cal C}_3}$  &  
 $8\,{\cal C}_3+\frac{144}{5}$
 & $L[\frac32]$ & 1\\ \hline
  & $\frac16 + \frac16 \sqrt{49 + 20\, {\cal C}_3}$  &  
  $8\,{\cal C}_3 + \frac{208}{5}-\frac{32}{5}\, \sqrt{49 + 20 \,{\cal C}_3} $
& $L[1]\smallotimes \{-\tfrac12\} $ & 1 \\ \hline
 & $\frac76 + \frac16 \sqrt{49 + 20\, {\cal C}_3}$   &   
 $8\,{\cal C}_3+\frac{88}{5}-\frac85\,\sqrt{49 + 20 \,{\cal C}_3} $ & $L[1]\smallotimes \{-\tfrac12\} $ & 1 \\ \hline
 &  $\frac{11}6 + \frac16 \sqrt{49 + 20\, {\cal C}_3}$ & 
 $8\,{\cal C}_3+\frac{88}{5}+\frac85\,\sqrt{49 + 20 \,{\cal C}_3} $ & $L[1]\smallotimes \{+\tfrac12\} $ & 1 \\ \hline
 & $\frac{17}6 + \frac16 \sqrt{49 + 20\, {\cal C}_3}$  &  
 $8\,{\cal C}_3 + \frac{208}{5} +\frac{32}{5}\, \sqrt{49 + 20 \,{\cal C}_3}  $
& $L[1]\smallotimes \{+\tfrac12\} $ & 1 \\ \hline
 & $-\frac{1}6 + \frac16 \sqrt{49 + 20\, {\cal C}_3}$  &  
 $8\,{\cal C}_3 + \frac{280}{5} -\frac{40}{5}\, \sqrt{49 + 20 \,{\cal C}_3}$   & $L[\frac12]\smallotimes \{-1\} $  & 1 \\ \hline
 & $\frac{3}2 + \frac16 \sqrt{9 + 20\, {\cal C}_3}$  &  $8\,{\cal C}_3$ & $L[\frac12]\smallotimes \{0\} $  & 2 \\ \hline
 & $\frac{19}6 + \frac16 \sqrt{49 + 20\, {\cal C}_3}$  &
 $8\,{\cal C}_3 + \frac{280}{5} +\frac{40}{5}\, \sqrt{49 + 20 \,{\cal C}_3}  $
& $L[\frac12]\smallotimes \{+1\} $  & 1 \\ \hline
\end{tabular}
\label{eq:spin21}
\end{equation}
where we also list their multiplicities and the supermultiplets to which they belong. Comparing to the previous results, we find that all the mass eigenvalues exhibited in \eqref{eq:spin21} fit into and fully span the list identified in~\cite{Yamagishi:1983ri,Ekhammar:2021gsg,Karlsson:2021oxd} (up to an overall normalisation factor in the definition of mass). On the other hand, the seemingly missing eigenvalue (of the operator $\Delta_2$ in the notation of \cite{Ekhammar:2021gsg}) is precisely taken care of by the non-trivial multiplicity in the penultimate line of \eqref{eq:spin21}.

Similarly, we may extract the scalar masses from \eqref{eq:spectrum1-app} as
\begin{equation}
\begin{tabular}{c|cccc}
 & $\Delta$ & $m^2$ & $L[J]$  & \#  \\ \hline\hline
$\phi$ 
& $-\frac{1}6 + \frac16 \sqrt{49 + 20\, {\cal C}_3}$  &  
   $8\,{\cal C}_3    +\frac{136}{5} -  8 \sqrt{49+20\,{\cal C}_3}$ 
 & $L[\frac12]\smallotimes \{-1\} $  & 1 \\ \hline
 & $\frac{3}2 + \frac16 \sqrt{9 + 20\, {\cal C}_3}$  &  
  $8\,{\cal C}_3 -\frac{144}{5}$ & $L[\frac12]\smallotimes \{0\} $  & 2 \\ \hline
 & $\frac{19}6 + \frac16 \sqrt{49 + 20\, {\cal C}_3}$  &
   $8\,{\cal C}_3     +\frac{136}{5}  + 8\sqrt{49+20\,{\cal C}_3}$
 & $L[\frac12]\smallotimes \{+1\} $  & 1 \\ \hline
& $ - \frac32 + \frac16\sqrt{81 + 20\, {\cal C}}$ & 
 $8\, {\cal C}_3 + \frac{648}{5} -  \frac{72}{5} \,\sqrt{ 81+ 20 {\cal C}_3}$ &  $L[0]\smallotimes \{-\tfrac32\} $ & 1 \\ \hline
&  $- \frac12 + \frac16\sqrt{81 + 20\, {\cal C}} $& 
 $8\,{\cal C}_3       +\frac{288}{5}   - \frac{48}{5} \, \sqrt{81+20\,{\cal C}_3} $   
&  $L[0]\smallotimes \{-\tfrac32\} $ & 1 \\ \hline
&  $\frac16 + \frac16\sqrt{1+20\,{\cal C}}$ & 
 $8\, {\cal C}_3     -\frac{32}{5} - \frac{32}{5}  \, \sqrt{1+20\,{\cal C}_3}$ 
& $L[0]\smallotimes \{-\tfrac12\} $ & 2 \\ \hline
&  $\frac76 + \frac16\sqrt{1+20\,{\cal C}}$ & 
 $8\,{\cal C}_3-\frac{152}{5} - \frac85\,\sqrt{1+20{\cal C}_3} $ & $L[0]\smallotimes \{-\tfrac12\} $ & 2 \\ \hline
&  $\frac{11}6 + \frac{1}6\sqrt{1+20\,{\cal C}}$ & 
 $8\,{\cal C}_3-\frac{152}{5} + \frac85\,\sqrt{1+20{\cal C}_3} $ & $L[0]\smallotimes \{+\tfrac12\} $ & 2 \\ \hline
&  $\frac{17}6 + \frac{1}6\sqrt{1+20\,{\cal C}}$ & 
 $8\,{\cal C}_3    -\frac{32}{5}  + \frac{32}{5}  \, \sqrt{1+20\,{\cal C}_3}$
& $L[0]\smallotimes \{+\tfrac12\} $ & 2 \\ \hline
& $  \frac72 + \frac16\sqrt{81 + 20\, {\cal C}}$ &
 $8\,{\cal C}_3   +\frac{288}{5}   + \frac{48 }{5}  \, \sqrt{81+20\,{\cal C}_3}$
&  $L[0]\smallotimes \{+\tfrac32\} $ & 1 \\ \hline
&  $\frac92 + \frac16\sqrt{81 + 20\, {\cal C}} $& 
 $8\, {\cal C}_3 + \frac{648}{5} +  \frac{72}{5} \,\sqrt{ 81+ 20 {\cal C}_3}$ &  $L[0]\smallotimes \{+\tfrac32\} $ & 1 \\ \hline
\end{tabular}
\label{eq:spin0}
\end{equation}
Again, all these eigenvalues fit into the list of eigenvalues identified in~\cite{Nilsson:1983ru,Duff:1986hr,Ekhammar:2021gsg} (up to an overall normalisation factor and shift in the definition of scalar mass). Just as before, the seemingly missing eigenvalues (of the operators $\Delta_3$, $\Delta_L$ in the notation of \cite{Ekhammar:2021gsg}) are precisely taken care of by the non-trivial multiplicities in the last column of \eqref{eq:spin0}. In that same notation of \cite{Ekhammar:2021gsg}, the eigenvalues of $\Delta_3$ pick a definite sign, fixing all the potential ambiguities.

Similar, one can extract the masses of all fields in the other representation towers \eqref{eq:spectrum2}, as well as in the lower representations \eqref{eq:degeneracies}. In the latter, the general pattern of \eqref{eq:spin21}, \eqref{eq:spin0}, degenerates and only some of the potential eigenvalues are realised, the explicit values follow from the multiplet structure together with \eqref{eq:S7Spectrum}.

\section{Degeneracies in low representations on the squashed $S^7$}
\label{app:Degeneracies}

In this appendix, we summarise the ${\cal N}=1$ supermultiplets in the Kaluza-Klein spectrum on the left-squashed $S^7$ which appear in $\USp{4} \times \SU{2}$ representations with small Dynkin labels, such that some of the generic structures \eqref{eq:spectrum1}, \eqref{eq:spectrum2} degenerate. In particular, in some of these representation, the values of the label $s$ do no longer combine into full ${\rm SL}(2)$ representations, such that in these cases we revert to the notation of \eqref{eq:not1}. The full list of these supermultiplets is given by
\begin{equation}
\label{eq:degeneracies}
	\begin{split}
		\left[1,q,1\right]_{q>1} &: \quad L[\tfrac32]\smallotimes [0]\; \oplus \; L[1]\smallotimes[\tfrac12] \;\oplus\; 
		L[\tfrac12]\smallotimes[1]\; \oplus \; L[0]\smallotimes[\tfrac32] \,, \\
		\left[0,q,0\right]_{q>1} &: \quad  L[\tfrac32]\smallotimes [0]\; \oplus  \; L[0]\smallotimes[\tfrac32] \,, \\
		\left[k,1,k\right]_{k>1} &: \quad  L[\tfrac32]\smallotimes [0]\; \oplus \; L[1]\smallotimes[\tfrac12] \;\oplus\; 
		L[\tfrac12]\smallotimes[\tfrac12\smallotimes\tfrac12]\; \oplus\; L[0]\smallotimes[\tfrac32] \; \oplus\; L[0]\smallotimes\{+\tfrac12\}
	 \,, \\
		\left[k,0,k\right]_{k>1} &: \quad L[\tfrac32]\smallotimes [0]\; \oplus \;  L[1]\smallotimes\{-\tfrac12\} \; \oplus \; L[\tfrac12]\smallotimes\{0,+1\}\; \oplus  \; L[0]\smallotimes[\tfrac32]\,, \\
		\left[1,1,1\right] &: \quad L[\tfrac32]\smallotimes [0]\; \oplus \; L[1]\smallotimes[\tfrac12] \;\oplus\; 
		L[\tfrac12]\smallotimes[1] \; \oplus\; L[0]\smallotimes\{-\tfrac32,+\tfrac12,+\tfrac32\}  \,, \\
		\left[1,0,1\right] &: \quad L[\tfrac32]\smallotimes [0]\; \oplus \;  L[1]\smallotimes\{-\tfrac12\} \; \oplus \;  L[\tfrac12]\smallotimes\{+1\}  \; \oplus\; L[0]\smallotimes\{-\tfrac32,+\tfrac32\} \,, \\
		\left[0,1,0\right] &: \quad L[\tfrac32]\smallotimes [0] \; \oplus\; L[0]\smallotimes\{-\tfrac32,+\tfrac12,+\tfrac32\} \,, \\
		\left[0,0,0\right] &: \quad   L[0]\smallotimes\{-\tfrac12,\tfrac32\}   \;\oplus\;  A_1[\tfrac32] \,, \\
		\left[k,1,k+2\right]_{k>0} &: \quad L[1]\smallotimes[\tfrac12] \;\oplus\;  L[\tfrac12]\smallotimes[1]\; \oplus \; L[0]\smallotimes[\tfrac12] \,, \\
		\left[k,0,k+2\right]_{k>0} &: \quad  L[1]\smallotimes\{+\tfrac12\} \; \oplus \; L[\tfrac12]\smallotimes\{-1,0\} \,, \\
		\left[0,q,2\right]_{q>1} &: \quad L[1]\smallotimes[\tfrac12] \;\oplus\;  L[\tfrac12]\smallotimes[1] \,, \\
		\left[0,1,2\right] &: \quad  L[1]\smallotimes[\tfrac12] \;\oplus\;   L[\tfrac12]\smallotimes\{-1,+1\}   \,, \\
		\left[0,0,2\right] &: \quad L[1]\smallotimes\{+\tfrac12\} \; \oplus \; A_1[\tfrac12] \,, \\
		\left[k+2,0,k\right]_{k>0} &: \quad  L[1]\smallotimes[\tfrac12] \;\oplus\;  L[\tfrac12]\smallotimes[1]\; \oplus \; L[0]\smallotimes\{+\tfrac12\} \,, \\
		\left[2,q,0\right]_{q>0} &: \quad  L[1]\smallotimes[\tfrac12] \;\oplus\;  L[\tfrac12]\smallotimes[1] \,, \\
		\left[2,0,0\right] &: \quad  L[1]\smallotimes[\tfrac12] \;\oplus\;   L[\tfrac12]\smallotimes\{+1\}  \; \oplus \; A_1[\tfrac12]  \,, \\
		\left[k,1,k+4\right] &: \quad L[\tfrac12]\smallotimes[0]\; \oplus \; L[0]\smallotimes\{-\tfrac12\} \,, \\
		\left[k,0,k+4\right] &: \quad L[0]\smallotimes\{-\tfrac12\} \,.
	\end{split}
\end{equation}
Note that for the $[0,0,0] L[0]\otimes\left\{-\tfrac12\right\}$ multiplet, our formula \eqref{eq:S7Spectrum} yields the value $\Delta = \frac13$, which lies below the unitary bound. This arises because KK spectroscopy strictly computes the mass eigenvalues in the AdS bulk, whereas their translation into conformal dimensions via $\Delta(\Delta-3) = m^2L^2$ allows for two solutions. For this multiplet, the other choice of solution of $\Delta$ gives the correct conformal dimension.\footnote{We thank Joel Karlsson for drawing our attention to this.}

This completes the full Kaluza-Klein spectrum.

\section{Spectrum of the right-squashed $S^7$}
\label{app:RS}

In this appendix, we summarize the entire spectrum of the right-squashed sphere, including the fermions. We extend what has been presented in Fig. \ref{fig:LeftRight12bos} as well as Fig. \ref{fig:LeftRight0bos} including the fermions, as well as present what happens for the $L[\frac32]$ and $L[1]$ long multiplets. 

%For the states of the $L[\tfrac32]$ multiplets, the transition between the right- and left-squashed sphere works as follows 

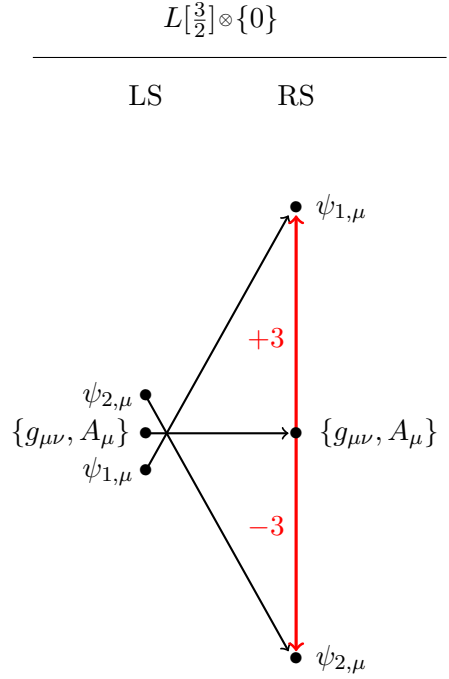
\begin{figure}[h!]
\begin{center}
\begin{tikzpicture}
	%Titre des zones 
	\node at (0,0) {$L[\frac32]\smallotimes\{0\}$};
	
	\node at (-1,-1)  {LS};
	\node at (1,-1)  {RS};
	
	\draw (-2.5,-0.5) -- (3,-0.5);

	\draw[thick] [->] (-1,-5) -- (0.9,-8.4) ;
	\draw[thick] [->] (-1,-5.5) -- (0.9,-5.5) ;
	\draw[thick] [->] (-1,-6) -- (0.9,-2.6) ;
	
	%\draw[dashed] (-1,-6) -- (1.25,-6) ;
	%\draw[dashed] (-1,-5) -- (1.25,-5) ;
	
	%\draw[blue,very thick] [->] (1.21,-6)--(1.21,-2);
	%\draw[blue,very thick] [->] (1.29,-5)--(1.29,-8.5);
	\draw[red,very thick] [->] (1,-5.5)--(1,-2.6);
	
	\draw[red,very thick] [->] (1,-5.5)--(1,-8.4);
		
	%\node at (1.6,-6.75) {\BL{$-\frac72$}};    
	%\node at (1.6,-4.25) {\BL{$+\frac72$}};    
	\node at (0.6,-6.75) {\EM{$-3$}};    
	\node at (0.6,-4.25) {\EM{$+3$}};    
		
	\node at (-2,-5.5) {$\{g_{\mu\nu},A_\mu\}$};
	\node at (-1.5,-6) {$\psi_{1,\mu}$};
	\node at (-1.5,-5) {$\psi_{2,\mu}$};
	\node at (2.1,-5.5) {$\{g_{\mu\nu},A_\mu\}$};
	\node at (1.6,-2.5) {$\psi_{1,\mu}$};
	\node at (1.6,-8.5) {$\psi_{2,\mu}$};
	
	\draw (-1,-5) node{$\bullet$} ;
	\draw (-1,-5.5) node{$\bullet$} ;
	\draw (-1,-6) node{$\bullet$} ;
	\draw (1,-2.5) node{$\bullet$} ;
	\draw (1,-8.5) node{$\bullet$} ;
	%\draw (1.25,-6) node{$\bullet$} ;
	%\draw (1.25,-5) node{$\bullet$} ;
	\draw (1,-5.5) node{$\bullet$} ;

\end{tikzpicture}
\end{center}
\caption{Shift patterns of the conformal dimensions within $L[\frac32]$, as we go from the left- to the right-squashed $S^7$. Here, $g_{\mu\nu}$ is a graviton, $A_{\mu}$ a vector and $\psi_{\mu}$ a gravitino.} \label{fig:LeftRight32}
\end{figure}

%For the states of the $L[\tfrac1]$ multiplets, the transition between the right- and left-squashed sphere works as follows 

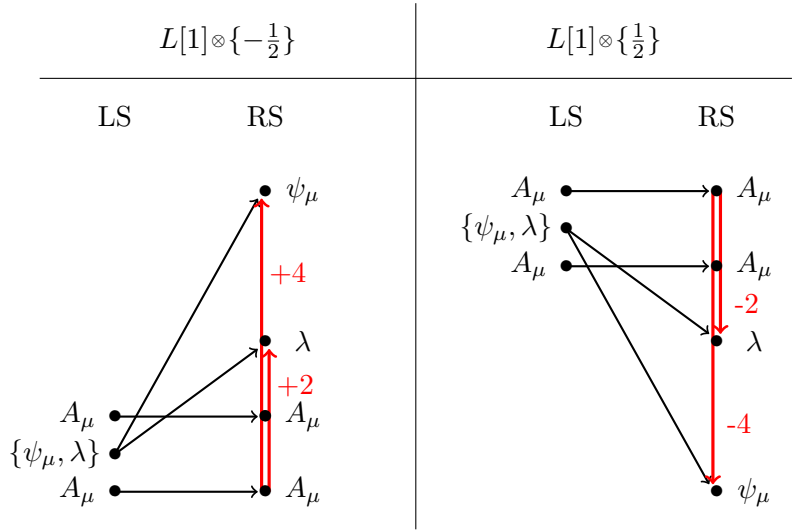
\begin{figure}[h!]
\begin{center}
\begin{tikzpicture}
    
    \draw (0,9.5) -- (10,9.5);   
    \draw (5,3.5) -- (5,10.5);
    
    %Left figure
    
    \node at (1,9)  {LS};
    \node at (3,9)  {RS};
    
    \node at (2.5,10) {$ L[1]\smallotimes \{-\frac12\}$}; 
    
    \node at (0.5,4) {$A_{\mu}$};
    \node at (0.5,5) {$A_{\mu}$};
    \node at (0.2,4.5) {$\{\psi_{\mu},\lambda\}$};
    \node at (3.5,4) {$A_{\mu}$};
    \node at (3.5,5) {$A_{\mu}$};
    \node at (3.5,6) {$\lambda$};
    \node at (3.5,8) {$\psi_{\mu}$};

    \draw[thick] [->] (1,4) -- (2.9,4) ;
    \draw[thick] [->] (1,5) -- (2.9,5) ;
    \draw[thick] [->] (1,4.5) -- (2.9,5.9) ;
    \draw[thick] [->] (1,4.5) -- (2.9,7.9) ;
    \draw[red,very thick] [->] (2.95,4) -- (2.95,7.9) ;
    \draw[red,very thick] [->] (3.05,4) -- (3.05,5.9) ;
    
    \node at (3.4,5.4) {\EM{+2}};
    \node at (3.3,6.9) {\EM{+4}};
    
    \draw (1,4) node{$\bullet$} ;
    \draw (1,4.5) node{$\bullet$} ;
    \draw (1,5) node{$\bullet$} ;
    \draw (3,4) node{$\bullet$} ;
    \draw (3,5) node{$\bullet$} ;
    \draw (3,5) node{$\bullet$} ;
    \draw (3,6) node{$\bullet$} ;
    \draw (3,8) node{$\bullet$} ;
    
    %Right figure

    \node at (7.5,10) {$ L[1]\smallotimes \{\frac12\}$}; 
    \node at (7,9)  {LS};
    \node at (9,9)  {RS};
    
    \node at (6.5,8) {$A_{\mu}$};
    \node at (6.5,7) {$A_{\mu}$};
    \node at (6.2,7.5) {$\{\psi_{\mu},\lambda\}$};
    \node at (9.5,8) {$A_{\mu}$};
    \node at (9.5,7) {$A_{\mu}$};
    \node at (9.5,6) {$\lambda$};
    \node at (9.5,4) {$\psi_{\mu}$};
    
    \draw[thick] [->] (7,8) -- (8.9,8) ;
    \draw[thick] [->] (7,7) -- (8.9,7) ;
    \draw[thick] [->] (7,7.5) -- (8.9,6.1) ;
    \draw[thick] [->] (7,7.5) -- (8.9,4.1) ;
    \draw[red,very thick] [->] (8.95,8) -- (8.95,4.1) ;
    \draw[red,very thick] [->] (9.05,8) -- (9.05,6.1) ;
    
    \node at (9.4,6.5) {\EM{-2}};
    \node at (9.3,4.9) {\EM{-4}};
    
    \draw (7,8) node{$\bullet$} ;
    \draw (7,7) node{$\bullet$} ;
    \draw (7,7.5) node{$\bullet$} ;
    \draw (9,8) node{$\bullet$} ;
    \draw (9,7) node{$\bullet$} ;
    \draw (9,6) node{$\bullet$} ;
    \draw (9,4) node{$\bullet$} ;
        
\end{tikzpicture}
\end{center}
\caption{Shift patterns of the conformal dimensions of the states within $L[1]$, as we go from the left- to the right-squashed $S^7$. Here, $A_{\mu}$ are vectors, $\psi_{\mu}$ is a gravitino, and $\lambda$ a fermion with spin $\frac12$.} \label{fig:LeftRight1}
\end{figure}

%For the states of the $L[\tfrac12]$ multiplets, the transition between the right- and left-squashed sphere works as follows 

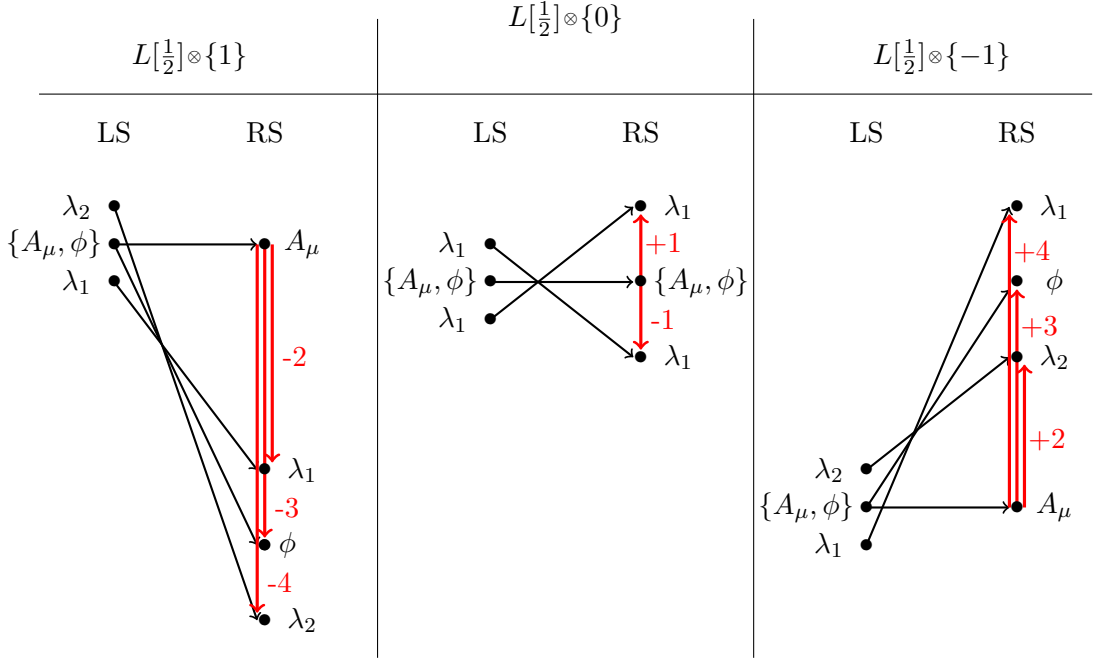
\begin{figure}[h!]
\begin{center}
\begin{tikzpicture}
    
    \draw (-2,9.5) -- (12,9.5);   
    \draw (2.5,2) -- (2.5,10.5);
    \draw (7.5,2) -- (7.5,10.5);    
	%Right figure
    \node at (10,10) {$ L[\tfrac12]\smallotimes \{-1\}$}; 
    
    \node at (9,9) {LS};
    \node at (11,9) {RS};
    
    \node at (8.2,4) {$\left\{A_{\mu},\phi\right\}$};
    \node at (11.5,7) {$\phi$};
    \node at (11.5,4) {$A_{\mu}$};
    \node at (11.5,8) {$\lambda_1$};
    \node at (11.5,6) {$\lambda_2$};
    \node at (8.5,3.5) {$\lambda_1$};
    \node at (8.5,4.5) {$\lambda_2$};
    
    \node at (11.3,6.4) {\EM{+3}};
    \node at (11.2,7.4) {\EM{+4}};
    \node at (11.4,4.9) {\EM{+2}};
    %\node at (11.6,6.4) {\BL{+3}};
    %\node at (11.5,7.4) {\BL{$+\frac92$}};
    %\node at (11.7,5.2) {\BL{$+\frac32$}};
    
    %\draw[dashed] (9,4) -- (11.3,4) ;
    %\draw[dashed] (9,4.5) -- (11.3,4.5) ;
    %\draw[dashed] (9,3.5) -- (11.3,3.5) ;
    \draw[thick] [->] (9,4) -- (10.9,6.9) ;
    \draw[thick] [->] (9,4) -- (10.9,4) ;
    \draw[thick] [->] (9,3.5) -- (10.9,8) ;
    \draw[thick] [->] (9,4.5) -- (10.9,6) ;
    \draw[red,very thick] [->] (11,4) -- (11,6.9) ;
    \draw[red,very thick] [->] (10.9,4) -- (10.9,7.9) ;
    \draw[red,very thick] [->] (11.1,4) -- (11.1,5.9) ;
    %\draw[blue,very thick] [->] (11.2,3.5) -- (11.2,7.9) ;
    %\draw[blue,very thick] [->] (11.3,4) -- (11.3,6.9) ;
    %\draw[blue,very thick] [->] (11.4,4.5) -- (11.4,5.9) ;
    
    \draw (9,4) node{$\bullet$} ;
    \draw (9,4.5) node{$\bullet$} ;
    \draw (9,3.5) node{$\bullet$} ;
    \draw (11,8) node{$\bullet$};
    \draw (11,6) node{$\bullet$};
    %\draw (11.3,4.5) node{$\bullet$};
    %\draw (11.3,4) node{$\bullet$};
    %\draw (11.3,3.5) node{$\bullet$};
    \draw (11,4) node{$\bullet$} ;
    \draw (11,7) node{$\bullet$} ;
       
%Left figure    
    
    \node at (0,10) {$L[\tfrac12] \smallotimes \{1 \}$ };
       
    \node at (-1,9) {LS};
    \node at (1,9) {RS};
    
    \node at (-1.8,7.5) {$\left\{A_{\mu},\phi\right\}$};
    \node at (-1.5,8) {$\lambda_2$};
    \node at (-1.5,7) {$\lambda_1$};
    \node at (1.5,7.5) {$A_{\mu}$};
    \node at (1.3,3.5) {$\phi$};
    \node at (1.5,2.5) {$\lambda_2$};
    \node at (1.5,4.5) {$\lambda_1$};

    \node at (1.3,4) {\EM{-3}};
    \node at (1.4,6) {\EM{-2}};
    \node at (1.2,3) {\EM{-4}};

    \draw[thick] [->] (-1,7.5) -- (0.9,7.5) ;
    \draw[thick] [->] (-1,7.5) -- (0.9,3.5) ;
    \draw[thick] [->] (-1,8) -- (0.9,2.5) ;
    \draw[thick] [->] (-1,7) -- (0.9,4.5) ;
    \draw[red,very thick] [->] (0.9,7.5) -- (0.9,2.6) ;
    \draw[red,very thick] [->] (1,7.5) -- (1,3.6) ;
    \draw[red,very thick] [->] (1.1,7.5) -- (1.1,4.6) ;
    
    \draw (-1,7.5) node{$\bullet$} ;
    \draw (-1,7) node{$\bullet$} ;
    \draw (-1,8) node{$\bullet$} ;    
    \draw (1,3.5) node{$\bullet$} ;
    \draw (1,7.5) node{$\bullet$} ;
    \draw (1,3.5) node{$\bullet$} ;
    \draw (1,4.5) node{$\bullet$} ;
    \draw (1,2.5) node{$\bullet$} ;
    
%middle figure

    \node at (5,10.5) {$ L[\tfrac12]\smallotimes \{0\}$};
    
    \node at (4,9) {LS};
    \node at (6,9) {RS};
    
    \node at (6.8,7) {$\left\{A_{\mu},\phi\right\}$};     
    \node at (3.2,7) {$\left\{A_{\mu},\phi\right\}$};
    \node at (3.5,6.5) {$\lambda_1$};
    \node at (3.5,7.5) {$\lambda_1$};
    \node at (6.5,6) {$\lambda_1$};
    \node at (6.5,8) {$\lambda_1$};
    
    \draw[thick] [->] (4,7.5) -- (5.9,6) ;
    \draw[thick] [->] (4,7) -- (5.9,7) ;
    \draw[thick] [->] (4,6.5) -- (5.9,8) ;
    \draw[red,very thick] [->] (6,7) -- (6,7.9) ;
    \draw[red,very thick] [->] (6,7) -- (6,6.1) ;
    
    \draw (4,7) node{$\bullet$} ;
    \draw (4,6.5) node{$\bullet$} ;
    \draw (4,7.5) node{$\bullet$} ;
    \draw (6,7) node{$\bullet$} ;
    \draw (6,8) node{$\bullet$} ;
    \draw (6,6) node{$\bullet$} ;

    \node at (6.3,7.5) {\EM{+1}};
    \node at (6.3,6.5) {\EM{-1}};
    
\end{tikzpicture}
\end{center}
\caption{Shift patterns of the conformal dimensions of the states within $L[\frac12]$, as we go from the left- to the right-squashed $S^7$. Here, $A_{\mu}$ is a vector, $\lambda$ are fermions with spin $\frac12$ and $\phi$ is a scalar field.} \label{fig:LeftRight12}
\end{figure}

%For the states of the $L[0]$ multiplets, the transition between the right- and left-squashed sphere works as follows 

\begin{figure}[h!]
\begin{center}
\begin{tikzpicture}
    \draw (2,2.5) -- (2,10.5);
    \draw (6,2.5) -- (6,10.5);
    \draw (10,2.5) -- (10,10.5);
    \draw (-2,9.5) -- (14,9.5);
    
        %Left left figure 
    \node at (0,10) {$L[0] \smallotimes \{-\tfrac32 \}$};
    \node at (4,10) {$L[0] \smallotimes \{-\tfrac12 \}$};
    \node at (8,10) {$L[0] \smallotimes \{\tfrac12 \}$};
    \node at (12,10) {$L[0] \smallotimes \{\tfrac32 \}$};
    
    \node at (-1,9) {LS};
    \node at (1,9) {RS};
    
    \node at (-1.4,3.5) {$\lambda$};
    \node at (-1.4,3) {$\phi_1$};
    \node at (-1.4,4) {$\phi_2$};
    \node at (1.4,3) {$\phi_1$};
    \node at (1.4,8) {$\lambda$};
    \node at (1.4,7) {$\phi_2$};
    
    \node at (1.4,5) {\EM{+4}};    
    \node at (1.3,7.5) {\EM{+5}};    
    
    \draw[thick] [->] (-1,3) -- (0.9,3) ;
    \draw[thick] [->] (-1,3.5) -- (0.9,7.9) ;
    \draw[thick] [->] (-1,4) -- (0.9,6.9) ;
    \draw[red,very thick] [->] (1.05,3) -- (1.05,6.9) ;
    \draw[red,very thick] [->] (0.95,3) -- (0.95,7.9) ;
    
    \draw (-1,3) node{$\bullet$} ;
    \draw (-1,3.5) node{$\bullet$} ;
    \draw (-1,4) node{$\bullet$} ;
    \draw (1,3) node{$\bullet$} ;
    \draw (1,7) node{$\bullet$} ;
    \draw (1,8) node{$\bullet$} ;
    
    % Left figure 
    
    \node at (3,9) {LS};
    \node at (5,9) {RS};
    \node at (2.6,6) {$\phi_2$};
    \node at (2.6,5.5) {$\lambda$};
    \node at (2.6,5) {$\phi_1$};
    \node at (5.4,6) {$\phi_2$};
    \node at (5.4,7) {$\lambda$};
    \node at (5.4,8) {$\phi_1$};
    
    \node at (5.3,7.4) {\EM{+2}};    
    \node at (5.4,6.4) {\EM{+1}};    
    
    \draw[thick] [->] (3,6) -- (4.9,6) ;
    \draw[thick] [->] (3,5) -- (4.9,7.9) ;
    \draw[thick] [->] (3,5.5) -- (4.9,6.9) ;
    \draw[red,very thick] [->] (5.05,6) -- (5.05,6.9) ;
    \draw[red,very thick] [->] (4.95,6) -- (4.95,7.9) ;
 
    \draw (3,5) node{$\bullet$} ;
    \draw (3,6) node{$\bullet$} ;
    \draw (3,5.5) node{$\bullet$} ;
    \draw (5,7) node{$\bullet$} ;
    \draw (5,8) node{$\bullet$} ;
    \draw (5,6) node{$\bullet$} ;
    
    %Right figure

    \node at (7,9) {LS};
    \node at (9,9) {RS};
    
    \node at (6.6,7) {$\phi_1$};
    \node at (6.6,8) {$\phi_2$};
    \node at (6.6,7.5) {$\lambda$};
    \node at (9.4,7) {$\phi_1$};
    \node at (9.4,5) {$\phi_2$};
    \node at (9.4,6) {$\lambda$};
    
    \node at (9.2,5.5) {\EM{-2}};
    \node at (9.3,6.6) {\EM{-1}};
    
    \draw[thick] [->] (7,8) -- (8.9,5.1) ;
    \draw[thick] [->] (7,7) -- (8.9,7) ;
    \draw[thick] [->] (7,7.5) -- (8.9,6.1) ;
    
    \draw[red,very thick] [->] (8.95,7) -- (8.95,5.1) ;
    \draw[red,very thick] [->] (9.05,7) -- (9.05,6.1) ;

    \draw (7,8) node{$\bullet$} ;
    \draw (7,7.5) node{$\bullet$} ;
    \draw (7,7) node{$\bullet$} ;
    \draw (9,7) node{$\bullet$} ;
    \draw (9,5) node{$\bullet$} ;    
    \draw (9,6) node{$\bullet$} ;
        
    %right right figure 
    \node at (11,9) {LS};
    \node at (13,9) {RS};
    
    \node at (10.6,8) {$\phi_2$};
    \node at (10.6,7) {$\phi_1$};
    \node at (10.6,7.5) {$\lambda$};
    \node at (13.4,8) {$\phi_2$};
    \node at (13.4,3) {$\lambda$};
    \node at (13.4,4) {$\phi_1$};
    
    \node at (13.4,6) {\EM{-4}};
    \node at (13.3,3.5) {\EM{-5}};
    
    \draw[thick] [->] (11,8) -- (12.9,8) ;
    \draw[thick] [->] (11,7) -- (12.9,4.1) ;
    \draw[thick] [->] (11,7.5) -- (12.9,3.1) ;
    \draw[red,very thick] [->] (12.95,8) -- (12.95,3.1) ; 
    \draw[red,very thick] [->] (13.05,8) -- (13.05,4.1) ; 
    
    \draw (11,8) node{$\bullet$} ;
    \draw (11,7) node{$\bullet$} ;
    \draw (11,7.5) node{$\bullet$} ;
    \draw (13,8) node{$\bullet$} ;
    \draw (13,4) node{$\bullet$} ;  
    \draw (13,3) node{$\bullet$} ;

\end{tikzpicture}
\end{center}
\caption{Shift patterns of the conformal dimensions of the states within $L[0]$, as we go from the left- to the right-squashed $S^7$. Here, $\lambda$ is a fermion with spin $\frac12$ and $\phi$ are scalar fields.} \label{fig:LeftRight0}
\end{figure}
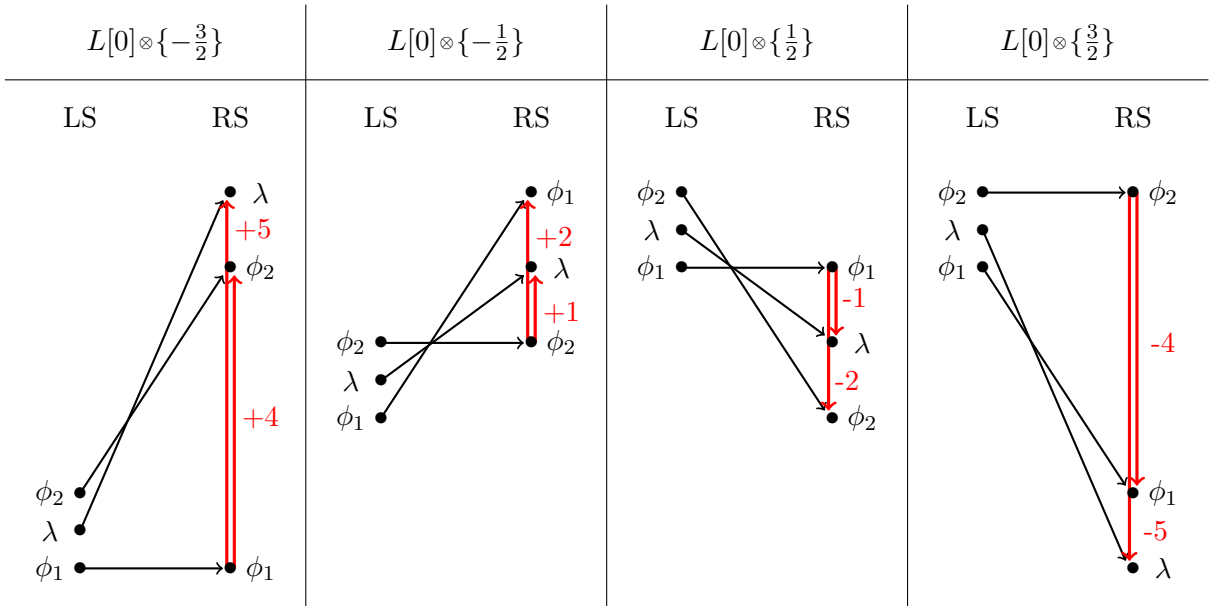

%\bibliographystyle{utphys}
%\bibliography{refs}

\clearpage

\providecommand{\href}[2]{#2}\begingroup\raggedright\endgroup

\end{document}